\PassOptionsToPackage{dvipsnames}{xcolor}

\documentclass[aps,prl,twocolumn,superscriptaddress,groupedaddress]{revtex4}  
\usepackage{graphicx}  
\usepackage{dcolumn}   
\usepackage{bm}        
\usepackage{amssymb}   
\usepackage{amsmath}   

\usepackage{hyperref}
\usepackage{tikz}
\usetikzlibrary{quantikz2}
\usepackage{mathtools}
\usepackage{placeins}
\usepackage{url}
\usepackage{algorithm}
\usepackage{algpseudocode} 

\newtheorem{definition}{Definition}
\newtheorem{proposition}{Proposition}
\newtheorem{theorem}{Theorem}

\begin{document}

\title{Quantum Metropolis–Hastings via Penalised Qubitized Walks:\\ Spectral Filtering and Circuit Implementation}

\author{Miguel Carrasco-Arango}\email{miguel.carrasco@bsc.es}\affiliation{Barcelona Supercomputing Center, 08034 Barcelona, Spain}
\author{Rosa M. Badia}\affiliation{Barcelona Supercomputing Center, 08034 Barcelona, Spain}
\author{Artur Garcia-Saez}\affiliation{Barcelona Supercomputing Center, 08034 Barcelona, Spain}\affiliation{Qilimanjaro Quantum Tech, 08019 Barcelona, Spain}
\begin{abstract}

The Metropolis-Hastings algorithm is a cornerstone of Markov Chain Monte Carlo methods, underpinning a wide range of applications in computational physics, Bayesian inference, and machine learning. Quantum variants of Metropolis-Hastings promise accelerated mixing through quantum walks, but their practical realisation remains challenging. In this work, we construct and simulate an explicit circuit level implementation of a quantum Metropolis-Hastings algorithm based on the framework introduced by Claudon \emph{et al.} (arXiv:2506.11576). We present the full quantum workflow to prepare a stationary distribution, including a number of modifications required to make the algorithm implementable in a realistic quantum circuit model. Our results demonstrate that these modifications are essential to recover the correct stationary behaviour and highlight both the potential and current limitations of quantum Metropolis-Hastings algorithms, which are expected to become practically relevant in the fault tolerant quantum computing regime.

\end{abstract}
\maketitle

\section{\label{sec:level1}Introduction}
Monte Carlo methods, and in particular Markov chain Monte Carlo (MCMC) methods, are a cornerstone of modern computational science. They are widely used in the simulation of stochastic systems, such as radiation transport \cite{RadiationTransport1, MonteCarloMethodsBook}, in the study of quantum many body systems, where they form a crucial component of machine learning approaches such as Neural Quantum States \cite{NQS1, NQS2}, and in Bayesian inference, where they are applied to a broad range of problems including inverse problems \cite{InverseProblems1, InverseProblems2}. However, despite their widespread success, the performance of classical MCMC methods is not uniform across all problem classes, and there exist important settings in which these methods become computationally inefficient.

The inefficiency of MCMC typically manifests as slow mixing in realistic high-dimensional settings. Slow mixing implies that a Markov chain requires a very large number of iterations to approximate the target distribution, which constitutes the main practical limitation of these methods. This behaviour is typically caused by structural features of the target distribution, such as high-dimensionality and poor scaling properties \cite{MCMCProblems1,MCMCProblems2}. In inverse problems, for instance, standard MCMC algorithms exhibit dimension-dependent mixing times that deteriorate under mesh refinement \cite{MCMCProblems3}. Another major obstacle is multimodality and the presence of large energy barriers: when the target distribution contains multiple well separated modes, strategies limited to local mechanisms can become trapped in a single region, leading to exponentially slow mixing in the worst case \cite{MCMCProblems4}. Since the mixing time scales inversely with the spectral gap of the Markov chain, poor designs based on local modifications on complex or high-dimensional targets can drastically reduce the gap, causing the required chain length to grow prohibitively large.

In his work \cite{Szegedy}, Szegedy introduced a general framework for quantizing algorithms based on classical random walks. In particular, he showed that any reversible Markov chain can be associated with a quantum walk whose spectral gap is quadratically larger than that of the original chain. Because the mixing time is governed by the inverse of the spectral gap \cite{markovchainslevin}, this result suggests a potential quadratic speedup for sampling tasks. Subsequent works have explored quantum versions of MCMC algorithms and established important theoretical foundations, but a key difficulty has remained: the intrinsic irreversibility introduced by rejected moves in Metropolis type algorithms, which complicates their direct implementation using quantum gates \cite{Lemieux,Temme,Montanaro}. Several proposed approaches address this issue by introducing auxiliary coin registers that encode acceptance or rejection decisions. While this allows the algorithm to remain unitary, the irreversibility is effectively shifted into controlled operations acting on the coin, which can involve computationally expensive reflections and may partially offset the expected quantum advantage. Moreover, since these constructions do not strictly follow Szegedy's quantization of Markov chains, their performance is typically analysed in a more heuristic manner.

A recent breakthrough was presented by Claudon \emph{et al.} \cite{Claudon}, who introduced a novel approach based on enlarging the state space and working on the dual (edge) representation of the proposal graph rather than on its vertices. Within this framework, they showed how a Metropolis-Hastings algorithm can be encoded into a qubitized quantum walk possessing a unique eigenstate with eigenvalue $1$ in the image of a specific partial isometry, which encodes the target stationary distribution.

In this work, we show that while the theoretical construction presented in \cite{Claudon} is mathematically sound, it cannot be directly implemented in a quantum circuit in a straightforward manner. In general, it is not possible to simultaneously apply a projection onto the $1$-eigenspace of the qubitized walk operator and a projection onto the range of the associated partial isometry, due to non-commutativity. We address this issue by introducing a penalised qubitized walk operator that lifts the degeneracy of the eigenvalue $1$ while preserving the desired stationary state. 

To assess the practical behaviour of the proposed construction, we design a workflow that enables an efficient circuit level implementation of the quantum Metropolis-Hastings algorithm and present the explicit quantum circuits required to realise it. We then perform classical simulations on high performance computing resources for two representative systems: a two-dimensional double well potential defined on a $4\times4$ grid and a four spin Ising chain. These examples serve as testbeds to analyse the performance and characteristic behaviour of the algorithm. In particular, they allow us to study how the algorithm behaves as the spectral gap of the underlying Markov chain narrows, to examine its effectiveness as a Gibbs state preparation method, and to investigate the properties of the spectral filter generated by quantum phase estimation when the available precision is insufficient to fully resolve the spectrum.

This article is organised as follows. In Section \ref{sec:level2}, we review the theoretical framework required for our study, including the basic theory of Markov chains, the classical Metropolis-Hastings algorithm, and the relevant aspects of the quantum construction introduced in \cite{Claudon}. In Section \ref{sec:level4}, we discuss the main difficulties that arise when attempting to directly implement Claudon's algorithm in a quantum circuit and present the workflow we have designed to overcome these obstacles. Finally, in Section \ref{sec:level5}, we present the numerical simulations for the double well potential and the Ising model and analyse the performance of the resulting implementation.

\section{Theoretical framework\label{sec:level2}}
\subsection{Classical Metropolis-Hastings}
The Metropolis-Hastings algorithm is a Markov chain Monte Carlo (MCMC) method used to sample from probability distributions, especially when the target distribution is intractable and cannot be sampled directly. The algorithm simulates a Markov process and produces a sample $\{X^{(n)}\}^N_{n=1}$, where the random variables $X^{(n)}\in E$ (the state space) are correlated and only approximately distributed according to the target distribution, which coincides with the stationary distribution of the Markov process $\pi$ \cite{firstcourseMCMC}.

Before introducing the Metropolis-Hastings algorithm in detail, we briefly recall here the basic notions of Markov chain theory that will be used throughout this work. In particular, we adopt the mathematical framework and terminology developed in \cite{Claudon}, which is well suited to the analysis and quantisation of Markov processes. For convenience and consistency with that framework, we recall the definitions of Markov kernels, irreducibility, stationarity, and ergodicity on finite state spaces.

\begin{definition}[\textbf{Markov kernel}]
    A Markov kernel on a finite state space $E$ is a function $P:E\times E\rightarrow [0,1]\subset\mathbb{R}$ such that 
    \begin{equation}
        P(x,y)\ge0 \; \forall x,y\in E,
    \end{equation} 
    and 
    \begin{equation}
        \sum_{y\in E} P(x,y)=1 \; \forall x \in E.
    \end{equation} 
    A Markov chain with kernel $P$ is said to be \textbf{reversible with respect to} a probability distribution $\pi$ if 
    \begin{equation}
        \pi(x)P(x,y)=\pi(y)P(y,x) , \; \forall x,y\in E.
    \end{equation}
\end{definition}

\begin{definition}[\textbf{Irreducible Markov chain}]
    A Markov chain with kernel $P$ on a finite state space $E$ is said to be \textbf{irreducible} if for all $x,y\in E$ there exists $t\in\mathbb{N}$ such that 
    \begin{equation}
        P^t(x,y)>0.
    \end{equation} 
\end{definition}

\begin{proposition}
    Let $P$ be a Markov kernel on a finite state space $E$ with a single non-degenerate eigenvalue on the complex unit circle, $\mathbb{S}^1=\{z\in \mathbb{C}     : |z|=1\}$. Then there exists a function $\pi:E\rightarrow[0,1]$ such that 
    \begin{equation}
        \forall x,y\in E, \; \lim_{t\to\infty} P^t(x,y)=\pi(y).
    \end{equation} 
\end{proposition}

\begin{definition}[\textbf{Stationary distribution}]
    Given a Markov kernel $P$ defined on a finite state space $E$, the probability vector $\pi$ defined as in the previous proposition is called the \textbf{stationary distribution} of $P$.
\end{definition}
\begin{definition}[\textbf{Ergodicity}]    
   A Markov chain on a finite state space is called \textbf{ergodic} if it is \textbf{irreducible} and has a single non-degenerate eigenvalue on the unit circle. 
\end{definition}

After introducing these concepts from Markov chain theory, we can now describe the Metropolis-Hastings algorithm. It is designed as a reversible ergodic Markov kernel that is naturally decomposed into two components: a proposal kernel $T$ that generates candidate moves, and an acceptance kernel $A$ that determines whether the proposed update is accepted or the state remains unchanged.

\begin{definition}[\textbf{Metropolis-Hastings kernel}]
    Given a probability distribution $\pi:E\rightarrow (0,1)$, one can construct an ergodic Markov kernel $P$ that is reversible with respect to $\pi$. This kernel is called the \textbf{Metropolis-Hastings kernel} and is defined as
    \begin{equation}
        P(x,y)=
        \begin{cases}
        T(x,y)A(x,y) \; &\text{if} \; x\neq y\\
        1-\sum_{z\in E}T(x,z)A(x,z) \; &\text{if} \; x=y
        \end{cases}
    \end{equation}
    where $T$ is the \textbf{proposal kernel}, satisfying $T(x,y)>0\implies T(y,x)>0$ and $T(x,x)=0$, and $A$ is the \textbf{acceptance kernel} with Metropolis choice,
    \begin{equation}
        A(x,y)=\min \left( 1,\frac{\pi(y)T(y,x)}{\pi(x)T(x,y)}\right),
    \end{equation} 
    for all $x,y\in E$ such that $T(x,y)>0$.
    The convergence properties of the Markov chain are governed by its \textbf{spectral gap}
    \begin{equation}
        \delta = 1 - \max_{\lambda \in \sigma(P)\setminus\{1\}} |\lambda|,
    \end{equation}
    defined as the difference between the leading eigenvalue $1$ and the second largest eigenvalue in modulus. For ergodic reversible chains, the mixing time scales inversely with this quantity, $\tau(\epsilon)=\mathcal{O}(1/\delta)$.

\end{definition}

\subsection{\label{sec:DualMH} Dual Metropolis-Hastings construction}
In order to overcome the irreversibility problem associated with the rejection of proposed moves, Claudon et al. \cite{Claudon} introduced an elegant idea consisting of enlarging the classical state space from $E$ to $E\times E$ and defining a Markov process on this extended (edge) state space. Accordingly, we work with the pairs $(x,y)\in E\times E$, where $x$ is denoted as the \textbf{tail} and $y$ as the \textbf{head}. We denote the set of directed edges of the proposal graph by 
\begin{equation}
    \mathcal{S} =\{(x,y)\in E\times E \;|\; T(x,y)>0\}.
\end{equation}

As before, we define a dual proposal kernel $\mathcal{T}$ and a dual acceptance kernel $\mathcal{A}$. In this construction, $\mathcal{T}$ always returns edges in $\mathcal{S}$ and updates only the tail of the edge.

\begin{definition}[\textbf{Dual Metropolis-Hastings}]
    Given a classical Metropolis-Hastings kernel $P$ that is reversible with respect to a probability distribution $\pi$, the \textbf{dual Metropolis-Hastings kernel} is defined as $\mathcal{P}=\mathcal{T}\mathcal{A}$, where the \textbf{dual proposal kernel} is
    \begin{equation}
        \mathcal{T}\left( (x,y),(z,t)\right)=\delta_x(z)T(x,t) ,
    \end{equation}
    for all $(x,y),(z,t)\in E\times E$, and the \textbf{dual acceptance kernel} is
    \begin{equation}
        \mathcal{A}\left((x,y),(z,t)\right)=
        \begin{cases}
            A(x,y), &\text{if } (z,t)=(y,x) \\
            1-A(x,y), &\text{if } (z,t)=(x,y) \\
            0, &\text{otherwise.}
        \end{cases}
    \end{equation}
    The kernel $\mathcal{P}$ acts on the first marginals in the same way as $P$.
\end{definition}

\begin{definition}[\textbf{Dual stationary distribution}]
    Given a classical Metropolis-Hastings kernel $P$ that is reversible with respect to a probability distribution $\pi$, the \textbf{dual stationary distribution} $\nu:E\times E\rightarrow[0,1]$ is defined as
    \begin{equation}
        \nu(x,y)=\pi(x)T(x,y) \quad\forall x,y\in E.
    \end{equation}
\end{definition}
The first marginal of $\nu$ is $\pi$. Under the above assumptions, both $\mathcal{T}$ and $\mathcal{A}$ are reversible with respect to $\nu$, and $\nu$ is a stationary distribution of $\mathcal{P}$. However, $\mathcal{P}$ is not necessarily reversible with respect to $\nu$. 

\subsection{Projected unitary encoding}

The mathematical framework described above provides an abstract characterization of the operators required to encode the Metropolis-Hastings dynamics as a quantum walk. To translate this description into an explicit quantum algorithm, one must specify how these operators are implemented as unitary transformations within a quantum circuit. In particular, this involves working with Hilbert spaces of different dimensions and embedding non square operators into larger unitary operators.

\begin{definition}[\textbf{Partial isometry}]
    Given two Hilbert spaces of different dimensionality $\mathcal{H}_1\cong\mathbb{C}^{n_1}$ and $\mathcal{H}_2\cong\mathbb{C}^{n_2}$, where $n_1<n_2$, with $n_1,n_2\in \mathbb{N}$. A \textbf{partial isometry} is a transformation 
    \begin{equation}
        \square:\mathcal{H}_1\rightarrow \mathcal{H}_2,
    \end{equation}
    where $\square^\dagger\square$ is the projection onto the support of $\square$.
\end{definition}

Since quantum computation is unitary, non-unitary matrices must be embedded into larger unitary operators in order to be implemented in a quantum circuit. This is achieved using the framework of projected unitary encodings introduced in \cite{Generalized_QSVT}.

\begin{definition}[\textbf{Projected Unitary Encoding}]
    Let $A$ be a complex matrix satisfying $\alpha \geq \|A\|_2$  (the spectral norm), let $U$ be a unitary operator and let $\square_L,\square_R$ be partial isometries. The triplet $(U, \square_L, \square_R)$ is called a \textbf{Projected Unitary Encoding (PUE)} of $A$. This encodes the matrix A up to a normalisation factor,  
    \begin{equation}
        \square_L^\dagger U\square_R=A/\alpha.
    \end{equation}
    The normalisation factor $\alpha$ ensures that $\|A/\alpha\|_2\le 1$, which is necessary for the existence of a unitary encoding. For notational simplicity, we will henceforth absorb this factor into $A$ and assume $\alpha=1$.

    If $\square_L=\square_R=\square$ and $U$ is symmetric, the pair $(U,\square)$ is called a \textbf{Symmetric Projected Unitary Encoding (SPUE)} of $A=\square^\dagger U \square$.
\end{definition}

\begin{definition}[\textbf{Qubitized walk operator}]
    Given a SPUE $(U,\square)$ of $A$, its \textbf{qubitized walk operator} is defined as
    \begin{equation}
        \mathcal{W}=\left( 2 \square \square^\dagger -Id \right)U.
    \end{equation}
    Each eigenvector $\ket{v_i}$ of $A$ with eigenvalue $v_i$ gives rise to the pair of eigenvalues $\lambda_i^{\pm}$ and corresponding eigenvectors $\ket{\lambda_i^\pm}$ of $\mathcal{W}$.
    If $v_i\in(-1,1)$, 
    \begin{equation}
        \lambda_i^\pm=e^{\pm i\cdot \gamma}
    \end{equation}
    with corresponding eigenvectors
    \begin{equation}
        \ket{\lambda_i^\pm}=\frac{1}{\sqrt{2}\sin(\gamma)}\left( e^{\pm i \cdot \gamma}-U\right)\square\ket{v_i},
    \end{equation}
    where $\gamma=\arccos(v_i)$. If $v_i=\pm 1$, then $\lambda_i=v_i$ and $\ket{\lambda_i}=\square\ket{v_i}$.
\end{definition}
In order to construct a qubitized walk operator, a SPUE is required. When only a non-symmetric PUE is available, this issue can be resolved by a standard hermitianization procedure.

\begin{definition}[\textbf{Hermitianization}]
    Given a PUE $(U, \square_L, \square_R)$ of $A$, its \textbf{hermitianization} is the tuple $(\overline{U},\overline{\square})$, where
    \begin{equation}
        \overline{U}=\left( \ket{0}\bra{0}\otimes U+ \ket{1}\bra{1}\otimes U^\dagger \right) \left( X\otimes Id \right),
    \end{equation}
    and
    \begin{equation}
        \overline{\square}=\ket{0}\bra{0}\otimes \square_L + \ket{1}\bra{1}\otimes \square_R.
    \end{equation}
    The pair $\left(\overline{U},\overline{\square}\right)$ is a SPUE of 
    \begin{equation}
        \overline{A}=\ket{0}\bra{1}\otimes A+ \ket{1}\bra{0}\otimes A^\dagger.
    \end{equation}
    The eigenvalues of $\overline{A}$ are the singular values of $A$ with both signs.
\end{definition}

\begin{definition}[\textbf{Discriminant}]
    Given a reversible ergodic Markov kernel $P$ with stationary distribution $\pi$, the \textbf{discriminant} of $P$ is defined as 
    \begin{equation}
        D(x,y)=\sqrt{\frac{\pi(x)}{\pi(y)}} P(x,y), \forall x,y\in E,
    \end{equation}
    or equivalently,
    \begin{equation}
        D= \text{diag}(\pi)^{1/2}\cdot P \cdot \text{diag}(\pi)^{-1/2}, 
    \end{equation}
    where diag$(\pi)$ is the diagonal matrix with the coordinates of the distribution $\pi$ in its diagonal. 
\end{definition}

\begin{proposition}
    Given a reversible ergodic kernel $P$ with stationary distribution $\pi$ and spectral gap $\delta$. If the tuple $(S,\square)$ is a SPUE of the discriminant $D$ of $P$, its qubitized walk operator is constructed as
    \begin{equation}
        \mathcal{W}=\left(2\square\square^\dagger-Id\right)S,
    \end{equation}
    where $S:\ket{x,y}\in\mathcal{H}_2\rightarrow S\ket{x,y}=\ket{y,x}\in\mathcal{H}_2$ is the swap operator. $\square\ket{\pi}$ is the only eigenvector of $\mathcal{W}$ in $\mathrm{Im}(\square)$. Moreover, the angular gap of $\mathcal{W}$ is quadratically larger than the one of $P$,
    \begin{equation}
        \begin{split}
            \Delta= \min\left\{ \theta\neq  0| e^{i\theta} \in \sigma(\mathcal{W}) \right\}& =\\
            =\arccos\left(\max\{\sigma(P) \setminus\{1\} \}\right)\in \;\Omega &\left(\sqrt{\delta}\right).
        \end{split}
    \end{equation}
    This yields a quadratic speed up respect to the classical mixing time \cite{discriminant}.
\end{proposition}

\section{\label{sec:level4} Algorithmic workflow}
With the relevant background established, we now summarize the main results of \cite{Claudon} that underpin the workflow considered in this section.

\begin{theorem}
    \label{teorema}
    Given a Metropolis-Hastings kernel $P$ with proposal kernel $T$, acceptance kernel $A$, stationary distribution $\pi$ and spectral gap $\delta$, it is possible to construct a unitary qubitized walk operator
    \begin{equation}
        \mathcal{W}=\left( 2\boxtimes \boxtimes^\dagger- Id\right) \left(X\otimes S\right),
    \end{equation}
    where $X$ is the $X$-Pauli gate and
    \begin{equation}
        \boxtimes=\ket{0}\bra{0} \otimes \square_* + \ket{1}\bra{1}\otimes\square.
    \end{equation}
    $\mathcal{W}$ has a unique eigenvector with eigenvalue $1$ in the image of $\boxtimes$, given by $\boxtimes\ket{+}\ket{\nu}\ket{0}$. Moreover, $\mathcal{W}$ has an angular gap in $\Omega\left( \sqrt{\delta} \right) $.
\end{theorem}

This theorem shows how to construct a qubitized walk operator that can be implemented as a quantum circuit, and whose unique $1-$eigenvector in the image of the partial isometry $\boxtimes$ encodes the dual stationary distribution. This state can be efficiently transformed into a quantum state encoding the stationary distribution $\pi$ of the classical Metropolis-Hastings kernel.

\begin{proposition}
    \label{decoding}
    Given the state $\boxtimes\ket{+}\ket{\nu}\ket{0}$ the following transformation allows one to sample from the stationary distribution $\pi$ by measuring the first register.
    \begin{equation}
    \begin{split}
        \big( Id\otimes O_T^\dagger\otimes Id\big)   \big( Id\otimes O^\dagger \big)& \big(\ket{0}\bra{0}\otimes S +\ket{1}\bra{1}\otimes\\
        \otimes Id\big) \boxtimes\ket{+}\ket{\nu}\ket{0}&=\ket{+}\ket{\pi,0}\ket{0},
    \end{split}
    \end{equation}
    where the oracles $O_T$ and $O$ are implemented as described in \cite{Claudon}. The first ket corresponds to an ancilla qubit, the second the first and second registers, and the third the remaining registers of the system.
\end{proposition}

These results establish the existence of a unitary operator $\mathcal{W}$ that encodes the Metropolis-Hastings dynamics in the form of a quantum walk. Restricted to the range of the partial isometry $\boxtimes$ the operator admits a unique eigenstate with eigenvalue $1$, which encodes the dual stationary distribution. A subsequent transformation maps this state to the target (edge) representation, from which the stationary distribution $\pi$ of the original Metropolis-Hastings kernel can be sampled. In the following subsections, we address the challenges involved in turning this abstract construction into an explicit workflow executable on a quantum circuit, and we motivate the design choices required to do so.

\subsection{Sequential filtering}

The results presented above reduce the implementation problem to the following task: filtering the $1-$eigenstates of $\mathcal{W}$, while simultaneously restricting to states in $\mathrm{Im}(\boxtimes)$. 
However, this task is more subtle than it may initially appear. In order to implement a projection onto the subspace
\begin{equation}
    \mathrm{Im}(\boxtimes)\cap \ker(\mathcal{W}-Id),
\end{equation}
where $\ker(\mathcal{W}-Id)$ denotes the eigenspace of
$\mathcal{W}$ corresponding to eigenvalue $1$, it is necessary that the corresponding projectors commute, namely
\begin{equation}
    \big[ \Pi_\boxtimes, \Pi_{+1}\big]=0.
\end{equation}

If this condition is not satisfied, successive applications of the two filters interfere with each other, and the action of the first projection may be partially or completely undone by the second.

Another crucial aspect concerns the implementation of the filtering procedure itself, since projections are non-unitary operations and cannot be directly realised within a quantum circuit. As a consequence, any filtering operation must be implemented indirectly using unitary dynamics together with measurement and classical postprocessing.

To this end, we introduce a filtering procedure based on Quantum Phase Estimation (QPE) \cite{QPE} to isolate the eigenstate of the qubitized walk operator $\mathcal{W}$ associated with eigenvalue $1$. Since the stationary state of $\mathcal{W}$ is characterised by a distinguished eigenphase $\theta=0$, a phase based filtering strategy provides a natural way to identify and extract this component of the spectrum.

\begin{figure*}[t]
    \centering
    \input{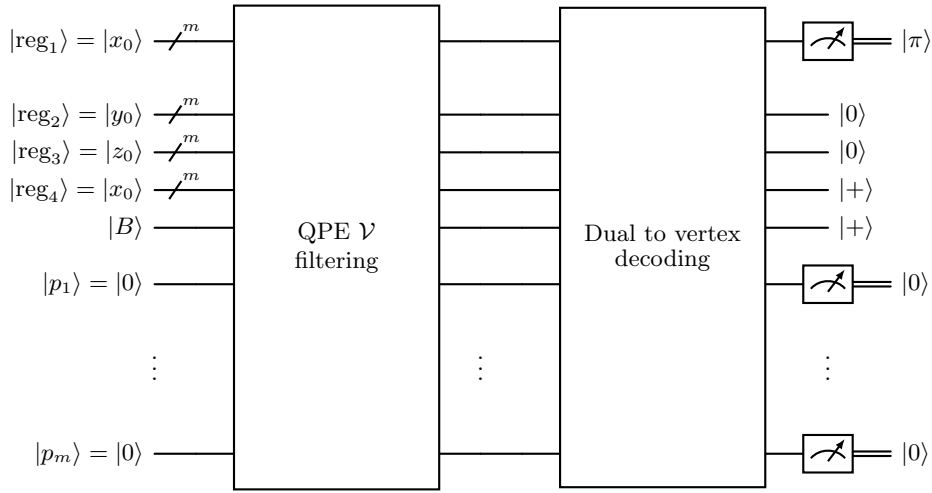}
    \caption{Scheme of the circuit implementing the Quantum Metropolis-Hastings. A QPE filtering is applied to an initial state encoded in 4 registers together with the required ancillas. The resulting state is then processed by a decoding transformation that maps the edge (dual) representation back to the vertex representation. Postselection on the $1-$eigenspace is performed by discarding all the states whose phase ancillas are not in the state $\ket{0}$. Finally, the stationary distribution is sampled by measuring the first register. }
    \label{workflow}
\end{figure*}

QPE provides a direct mechanism to access this phase information by coherently estimating the eigenphases of $\mathcal{W}$. By resolving the eigenphase to sufficient precision and postselecting on outcomes close to zero, one can effectively isolate the desired eigenstate. This approach aligns naturally with the structure of the qubitized walk operator and leads to a transparent filtering procedure whose resource requirements can be analysed directly in terms of phase resolution.

Alternative filtering strategies based on polynomial transformations of the spectrum, such as those enabled by quantum singular value transformation (QSVT) techniques \cite{Unification}, are in principle applicable. These methods operate by applying bounded polynomial functions to a suitable Hermitian signal derived from the unitary. However, in the present setting the target eigenstate corresponds to an eigenphase $\theta=0$, which maps to an eigenvalue located at the endpoint of the spectrum of the associated Hermitian signal operator. Approximating a sharp filter near such an endpoint would require high degree polynomial transformations, leading to significantly increased circuit complexity. For this reason, and to keep the construction aligned with the phase based nature of the problem, we restrict attention to the QPE based filtering approach and leave the exploration of QSVT based alternatives for future work.

We now describe how the QPE algorithm \cite{QPE} implements the filtering operation discussed above. Given a unitary operator $U$ with an eigenvector $\ket{\lambda}$ satisfying
$U\ket{\lambda} = e^{i2\pi\phi}\ket{\lambda}$, QPE produces a binary estimate of the phase $\phi$,
\begin{equation}
    \ket{\lambda}\ket{0}\xrightarrow{\text{QPE}}\ket{\lambda}\ket{\phi}.
\end{equation}
An arbitrary state can be decomposed in the eigenbasis of a unitary operator as
\begin{equation}
    \ket{\psi}=\sum_{j\geq 0}\alpha_j\ket{\lambda_j},
\end{equation}
where $\ket{\lambda_j}$ denotes an eigenstate of the operator with associated eigenphase $\phi_j$. Applying the QPE algorithm with $m$ ancilla qubits to the state $\ket{\psi}$ yields
\begin{equation}
    \ket{\psi_f}=\sum_{j\geq 0}\frac{\sin(2^{m-1}\phi_j)}{2^m \sin(\phi_j/2)}\alpha_j\ket{\lambda_j}\ket{\phi_j}.
\end{equation}
The amplitude prefactor defines the spectral response of the QPE filter as a function of the eigenphase $\phi_j$.

Postselecting on the phase register being measured in the state $\ket{0}$ therefore implements an approximate projection onto the $1-$eigenspace of $\mathcal{W}$. In the ideal limit of infinite phase resolution, this procedure
retains the eigenstate with eigenvalue $1$ with probability $\alpha_0^2$. However, when a finite number of ancilla qubits is used, the resulting filter is not exact. Instead, it realises a Dirichlet kernel \cite{DirichletKernel},
corresponding to a sharply peaked response at phase $0$ with small side lobes. These side lobes allow eigenstates with nearby phases to pass the filter with a probability that depends on the number of ancillas used:
\begin{equation}
    Pr(\phi_j)=\alpha_j^2\frac{\sin^2(2^{m-1}\phi_j)}{2^{2m} \sin^2(\phi_j/2)}.
\end{equation}

Therefore, since the Dirichlet kernel has its first zero at 
$|\phi_j| \approx 2\pi/2^m$, eigenphases smaller than this scale are not 
effectively suppressed by the filter. In particular, to isolate the 
$1$-eigenstate from the first excited mode with eigenphase $\phi_2$, 
we must require
\[
\frac{2\pi}{2^m} \lesssim |\phi_2|.
\]
Moreover, for a reversible Metropolis-Hastings chain with classical 
spectral gap $\delta = 1-\lambda_2$, the corresponding qubitized walk 
has eigenphases satisfying
\[
|\phi_2| = \Theta(\sqrt{\delta}).
\]
Hence,
\[
\frac{2\pi}{2^m} \lesssim \sqrt{\delta},
\qquad\implies\qquad
m \gtrsim \log_2\!\left(\frac{2\pi}{\sqrt{\delta}}\right).
\]

At this point, the obstruction identified above can be made explicit at the level of the circuit implementation. The QPE based filtering procedure requires multiple controlled applications of the walk operator
$\mathcal{W}$, corresponding to the powers of $\mathcal{W}$ appearing in the phase estimation circuit. Consequently, any constraint imposed during the filtering process must be preserved under repeated applications of $\mathcal{W}$.

In particular, enforcing the restriction to $\mathrm{Im}(\boxtimes)$ in conjunction with the phase based filter implicitly requires that this subspace be invariant under the action of $\mathcal{W}$, which is equivalent to the commutation condition
\begin{equation}
    [\Pi_\boxtimes, \mathcal{W}] = 0,
\end{equation}
and hence implies $[\Pi_\boxtimes, \Pi_{+1}] = 0$. An analogous requirement arises for QSVT based filtering schemes, which also rely on polynomial functions of $\mathcal{W}$ and therefore involve repeated applications of the walk operator.

However, in general the partial isometry $\boxtimes$ does not commute with $\mathcal{W}$. The action of the walk operator mixes states inside and outside $\mathrm{Im}(\boxtimes)$, so this subspace is not invariant under
$\mathcal{W}$. As a result, the restriction to $\mathrm{Im}(\boxtimes)$ is disrupted during the execution of the filtering circuit, and the naive combination of the two filtering operations fails. This necessitates a different strategy, which we develop in the following.

\subsection{Breaking the eigenvalue degeneracy}
\begin{figure*}[t]
    \centering
    \includegraphics[width=1\linewidth]{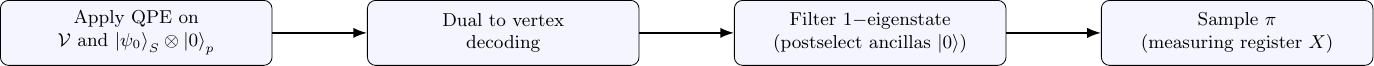}
    \caption{Flowchart of the quantum Metropolis-Hastings sampling procedure based on qubitization and Quantum Phase Estimation. The circuit is initialized in the state $\ket{\psi_0}_S\otimes \ket{0}_p$, after which the penalised qubitised walk operator $\mathcal{V}$ is constructed. Quantum Phase Estimation of $\mathcal{V}$ coherently lifts the components of $\ket{\psi_0}_S$ to the eigenbasis of $\mathcal{V}$, encoding the corresponding eigenphases in the phase register. A final transformation maps the resulting state to the target (edge) representation. The phase register is then postselected onto the $\ket{0}$ subspace, corresponding to the $1-$eigenspace of $\mathcal{V}$; all other branches are discarded. Finally, the first register of the system is measured to obtain a sample.}
    \label{flowchart}
\end{figure*}

Assuming that the spectral filter successfully extracts eigenstates of $\mathcal{W}$ with eigenvalue $1$, it remains to distinguish the unique eigenstate that lies in $\mathrm{Im}(\boxtimes)$ from the remaining ones. Indeed, while this eigenstate is unique within $\mathrm{Im}(\boxtimes)$, the eigenvalue $1$ is generally degenerate on the full Hilbert space.

A natural strategy to resolve this degeneracy is to modify the spectrum of $\mathcal{W}$ in such a way that the desired eigenstate remains unchanged, while all other eigenstates with eigenvalue $1$ acquire a non trivial phase and are thus shifted away from phase $0$.

This is achieved by defining a penalised qubitized walk operator,
\begin{equation}
    \mathcal{V}=\left(\Pi_\boxtimes+e^{i\varphi}\left(Id-\Pi_\boxtimes\right)\right)\mathcal{W}.
\end{equation}
By construction, this phase kickback acts trivially on states in $\mathrm{Im}(\boxtimes)$, while inducing a rotation on the unit circle for eigenstates outside this subspace. As a result, the global degeneracy of the eigenvalue $1$ is lifted, while the target eigenstate is preserved.

Due to the finite resolution of the spectral filter, it is in principle possible that some rotated eigenstates acquire phases sufficiently close to zero to partially pass the filter. In practice, this issue can be mitigated by repeating the procedure with different values of the penalty phase $\varphi$ and identifying the unique state that remains invariant under all such choices.

\subsection{Full algorithmic workflow}

We now present the workflow that results from the design choices introduced in the previous subsections. While the construction of the underlying qubitized walk operator follows the theoretical framework of \cite{Claudon}, the workflow described here provides a concrete circuit level realisation that integrates spectral filtering, symmetry breaking, and state preparation into a single executable procedure.

In the circuit implementation, we work with the set of qubits used to implement the transformations described in \cite{Claudon}, which we refer to as the \textbf{system qubits}. In addition, we introduce a set of \textbf{phase register ancillas} used to store the phase in the QPE based filter, as well as a \textbf{penalty flag qubit} that enables the conditional phase kick required for the penalised qubitized walk. 

The system qubits are organised into four main registers: $x$, encoding the tail vertex; $y$, encoding the head vertex; $z$, corresponding to the proposal register; and  $x_c$, which stores a copy of $x$ used in uncomputation. Each of these registers consists of $n_{\mathrm{sys}}=\ulcorner\log_2 |E|\urcorner$ qubits, where $|E|$ is the size of the classical state space. In addition, we use a single ancilla qubit $b$ to control the application of the partial isometries $\square$ or $\square_\star$, along with further ancillas required for the implementation of the oracles. 

In \cite{Claudon}, it is claimed that the number of qubits required to encode the core walk construction scales as $4n_{\mathrm{sys}}+3$, corresponding to the registers needed to represent the system state together with a small number of control ancillas. This count is given at the level of the abstract walk construction and assumes oracle access to the proposal and acceptance operations.

\begin{algorithm}[h]
\caption{Quantum Metropolis-Hastings workflow}
\label{alg:workflow}
\begin{algorithmic}[1]
\Require Proposal graph $E$, acceptance ratios $A(x,y)$ for neighbouring vertices,
penalty phase $\varphi$, QPE precision $m$
\Ensure Samples from the stationary distribution $\pi$
\State Construct the oracle operators $O_\mathcal{A}$, $O_T$ encoding the proposal kernel and acceptance rule
\State Construct the partial isometries $\square$, $\square_\star$ and the operator $\boxtimes$
\State Construct the penalised qubitized walk operator
\[
\mathcal{V} = \left(\Pi_\boxtimes + e^{i\varphi}(Id-\Pi_\boxtimes)\right)\mathcal{W}
\]
\State Initialise the system qubits in a seed state $\boxtimes\ket{\psi} \in \mathrm{Im}(\boxtimes)$
\State Initialise the phase register ancillas and penalty flag qubit in the state $\ket{0}$
\State Apply Quantum Phase Estimation on $\mathcal{V}$ with precision $m$
\State Apply the dual to vertex decoding transformation
\State Measure the phase register and postselect on outcome $\ket{0}$
\State Measure the register $x$ to obtain a sample from the stationary distribution $\pi$
\end{algorithmic}
\end{algorithm}

In the present workflow, the number of qubits required to encode the system registers remains $4n_{\mathrm{sys}}$, but additional resources are required to support the filtering and symmetry breaking mechanisms introduced above. These include the penalty flag qubit, the phase register ancillas used in the QPE based filter, as well as the ancillas required to explicitly implement the oracle constructions. The latter depend on the structure and topology of the proposal graph $E$ and therefore cannot be specified independently of a concrete instance.

Overall, the total qubit complexity of the workflow scales as
\begin{equation}
\mathcal{O}\!\left(4\ulcorner\log_2 |E| \urcorner+ m + n_{\mathrm{anc}}\right),
\end{equation}
where $m$ denotes the number of phase register ancillas and $n_{\mathrm{anc}}$ accounts for the ancillary qubits required by the oracle implementations. The explicit circuit constructions underlying this implementation are provided in Appendix \ref{app:circuits}.

The scheme of the quantum circuit implementation is illustrated in FIG. \ref{workflow}. For clarity, the complete procedure is summarised in the flowchart shown in FIG. \ref{flowchart} and  described in Algorithm \ref{alg:workflow}. The system is initialised in a state that ideally has a large overlap with the target $1-$eigenstate of the penalised walk operator. In practice, most of the time the only information available about the target state is that it lies in the range of $\boxtimes$. We therefore initialise the system in a seed state of the form $\boxtimes\ket{\psi}\in \mathrm{Im}(\boxtimes)$, where $\ket{\psi}$ is chosen as a uniform superposition over the directed edges of the proposal graph.

Next, QPE is applied using the penalised qubitized walk operator $\mathcal{V}$. After performing the dual to vertex decoding described in Proposition \ref{decoding}, the qubits are measured, and only those runs in which the phase register ancillas are in the state $\ket{0}$ are retained. Finally, measuring the register $x$ yields samples distributed according to the target stationary distribution $\pi$.

\section{\label{sec:level5} Simulations}

The quantum Metropolis-Hastings framework described above is formulated in terms of explicit quantum circuits, involving Quantum Phase Estimation, controlled reflections, and oracle constructions whose depth grows with the desired spectral precision. While these circuits are well defined within the fault tolerant quantum computing paradigm, their coherence and depth requirements place them beyond the capabilities of current noisy intermediate scale quantum (NISQ) devices.

For this reason, all results reported in this section are obtained via classical simulation of the corresponding quantum circuits. The simulations were performed on the MareNostrum 5 supercomputer, using a single node of the accelerated partition. Circuit construction and execution were carried out using the \texttt{Qiskit} framework, with statevector simulations performed via the \texttt{Qiskit Aer} module. This allows us to analyse the algorithm in a fully controlled setting and directly compare the prepared distributions with analytically known stationary states.

Two complementary models are studied in detail. We begin with a bidimensional double well defined on a $4\times4$ grid. Because the proposal geometry is spatial and visually intuitive, this example provides a transparent setting in which the action of the Metropolis dynamics can be directly interpreted in terms of movement across an energy landscape. In particular, it serves as a minimal multimodal test case that makes the role of the penalised walk operator especially clear. This first example therefore illustrates the structural importance of symmetry breaking in the quantum construction.

We then turn to the Ising model, a paradigmatic and widely used benchmark in statistical physics and Monte Carlo studies. Unlike the double well, the Ising model is naturally parametrised by the inverse temperature $\beta$, allowing us to continuously tune the spectral gap of the underlying Metropolis kernel. This makes it possible to systematically investigate how the performance of the quantum filtering procedure depends on problem difficulty, and to analyse in detail the transition between the fully resolved and under resolved regimes.

For the Ising simulations, instead of employing the standard QPE circuit with multiple phase register ancillas, we adopt the single ancilla iterative phase estimation technique of \cite{EfficientFactorization}. In this approach, only one phase qubit is required, and precision is increased through repeated controlled applications of the walk operator combined with sequential measurements. Each round refines the phase estimate, effectively trading circuit width for repetition depth. This implementation significantly reduces the ancillary qubit requirements and allows us to explicitly study how the prepared distribution converges as the phase resolution is progressively increased. 

Although realised semiclassically, this construction induces the same phase distribution, and therefore the same effective spectral filter, as the standard coherent QPE circuit (see Appendix \ref{app:semiclassical QPE filter} for circuit construction). In practice, this formulation enables us to reach higher phase precision in simulation without increasing the size of the ancilla register, providing a controlled way to analyse how resolution affects the prepared distribution.

Although the system sizes considered here are necessarily small due to the exponential cost of classical statevector simulation, we are able to simulate circuits involving at most $27$ qubits in total when implementing the fully coherent version of the algorithm, corresponding to the circuit that would be executed on a quantum computer. This constraint limits the maximal phase estimation precision that can be explored and therefore restricts the regime in which the spectral filtering procedure can be fully resolved. By contrast, the semiclassical single ancilla implementation reduces the number of simultaneously required qubits to $22$, while allowing us to reach higher phase precision within the same simulation budget. These system sizes are nevertheless sufficient to validate the correctness of the construction, to study its qualitative spectral behaviour, and to understand how finite phase resolution shapes the prepared distributions. Our objective is therefore not to demonstrate a quantum advantage, but to provide a complete, end to end implementation of quantum Metropolis-Hastings that can be rigorously analysed in a controlled computational environment.

Throughout the simulations we fix the penalisation phase of the walk operator to $\varphi = 1.0472$ radians. This value was chosen heuristically, as it was observed to provide stable convergence and effective spectral separation in the explored parameter regimes. A more systematic analysis of the optimal penalisation phase and its dependence on the spectral properties of the underlying kernel is left for future work.

\subsection{Bidimensional double well}
\subsubsection{Model and target distribution}

\begin{figure}[h]
    \centering
    \includegraphics[width=1\linewidth]{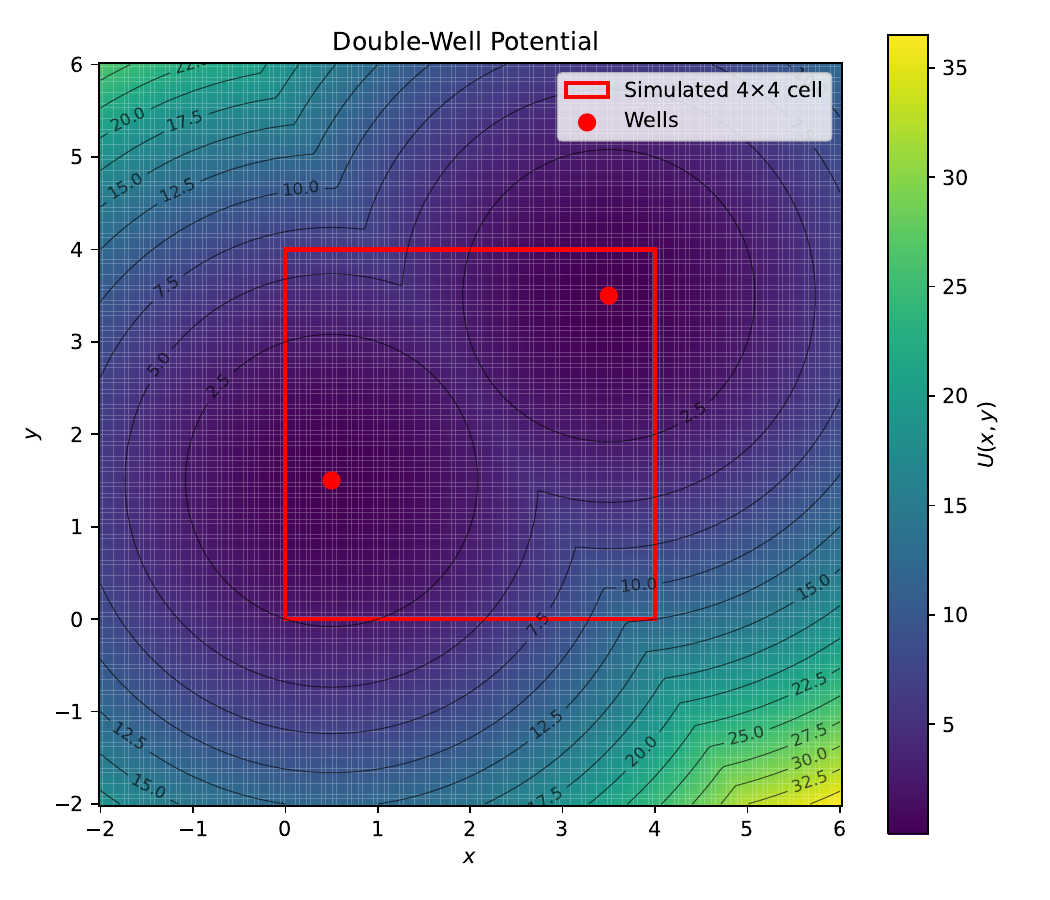}
    \caption{Visualization of the double well energy landscape. The figure shows a two-dimensional colourmap of the potential defining the target Boltzmann distribution. Two quadratic wells are located at $(0,1)$ and $(3,3)$. The red square highlights the fundamental $4\times4$ domain on which the Markov chain and its quantum implementation are defined. The extended domain is shown only for visualization purposes.}
    \label{fig:potential}
\end{figure}

To illustrate the behaviour of the quantum Metropolis-Hastings algorithm in the presence of multimodality, we consider a simple but representative double well energy landscape defined on a finite two-dimensional lattice (see Fig. \ref{fig:potential}). The figure provides a continuous visualization of the underlying landscape for geometric intuition, while the simulations themselves operate on a deliberately coarse $4\times4$ discretization. Despite its small size, this model already captures the essential difficulty faced by classical Markov chain Monte Carlo methods when sampling from distributions with multiple separated modes.

\paragraph{State space and proposal geometry.}
The system is defined on a $4\times4$ grid, so that the state space is
\begin{equation}
    E=\left\{(i,j)\;|\; i,j\in\{0,1,2,3\}\right\},
\end{equation}
consisting of only $|E|=16$ states. In order to keep the proposal kernel simple, uniform, and symmetric across all states, we endow this grid with periodic nearest neighbour connectivity and interpret it as a discrete torus. Each state $(i,j)$ therefore has four neighbours,
\begin{equation}
    \mathcal{N}(i,j)
    =
    \{(i\pm1,j),(i,j\pm1)\ \mathrm{mod}\ 4\}.
\end{equation}
This choice avoids boundary effects, ensures that the proposal distribution has constant degree, and prevents pathological behaviour such as walkers becoming effectively trapped near the edges of the domain. From an implementation perspective, it also leads to particularly simple and symmetric quantum proposal oracles.

\paragraph{Double well energy landscape.}
While the proposal geometry is taken to be periodic, the energy function itself is defined directly on the fundamental $4\times4$ cell and is not made periodic. We deliberately adopt this non periodic energy landscape in order to keep the model as simple and transparent as possible while retaining genuine multimodality.

The potential energy is defined as
\begin{equation}
    U(i,j)
    =
    \min\!\left(
        i^2+(j-1)^2,\;
        (i-3)^2+(j-3)^2
    \right),
\end{equation}
which introduces two competing quadratic wells with minima at $(0,1)$ and $(3,3)$. Despite the small state space, this energy landscape already exhibits a clear separation between metastable regions, making it a minimal but nontrivial test case for multimodal sampling.

The target distribution is chosen to be the Boltzmann (Gibbs) distribution associated with this potential,
\begin{equation}
    \pi(i,j)
    =
    \frac{1}{Z}\exp\!\left(-\frac{U(i,j)}{T}\right),
\end{equation}
where we fix the temperature to $T=1$ throughout, and
\begin{equation}
    Z=\sum_{i,j}\exp\!\left(-\frac{U(i,j)}{T}\right)
\end{equation}
is the partition function. The resulting stationary distribution is explicitly bimodal, with probability mass concentrated around the two wells.

\paragraph{Connection to optimization and multimodality.}
At low temperature, sampling from $\pi$ becomes increasingly concentrated near the minima of $U$. In this regime, the problem can be viewed both as a sampling task and as a discrete optimization problem, in which the objective is to identify and explore low energy configurations.

Multimodal energy landscapes of this type are well known to be challenging for classical Metropolis-Hastings algorithms with local proposals, which may remain trapped for long times in metastable regions separated by energy barriers. Similar difficulties arise in classical optimization, where escaping local minima typically requires nonlocal moves or carefully tuned annealing schedules. A wide range of quantum algorithms have been proposed to address such challenges, including quantum annealing and adiabatic approaches.

From this perspective, quantum Metropolis-Hastings algorithms can be viewed as a complementary framework, aimed not at directly locating a single optimal configuration, but at efficiently preparing and sampling from low temperature Gibbs states on structured energy landscapes. Given the $27$ qubit resource limit of our simulations, the filtering step remains in a finite precision (under resolved) regime. The simple model studied here therefore serves as a controlled testbed for analysing how quantum enhanced sampling techniques behave under realistic spectral resolution constraints in the presence of multimodality.

\paragraph{Proposal and acceptance kernels.}
The proposal kernel is chosen to be uniform over the four nearest neighbours on the torus,
\begin{equation}
    T(x,y)=
    \begin{cases}
        \frac{1}{4}, & y\in\mathcal{N}(x),\\[4pt]
        0, & \text{otherwise},
    \end{cases}
\end{equation}
which ensures symmetry and simplifies the quantum implementation. The Metropolis-Hastings acceptance rule is defined as
\begin{equation}
    A(x,y)
    =
    \min\!\left\{1,\; \frac{\pi(y)}{\pi(x)}\right\}
    =
    \min\!\left\{1,\; e^{-[U(y)-U(x)]}\right\},
    \label{eq:MH acceptance kernel}
\end{equation}
guaranteeing that $\pi$ is the stationary distribution of the resulting Markov chain despite the decoupling between proposal geometry and energy landscape.

Details of the quantum circuits implementing the proposal and acceptance oracles, as well as the quantitative metrics used to analyse the simulation results, are provided in Appendix \ref{app:circuits}.

\subsubsection{Results and analysis}
\label{sec: results double well}
We begin by analysing how the quality of the distribution prepared by the penalised quantum Metropolis-Hastings algorithm depends on the number $m$ of precision qubits used in the QPE filter. The
parameter $m$ controls the spectral resolution of the filtering step and thus directly affects how accurately the stationary distribution can be reproduced. 

\begin{table}[h]
    \centering
    \begin{tabular}{c c c c c}
        \hline
        $m$ & $F(p_X,\pi)$ & $d_{\mathrm{TV}}(p_X,\pi)$ & $P_{\mathrm{basin}}(p_X)$ & accept.\ prob. \\
        \hline
        1 & 0.8248 & $2.88\times 10^{-1}$ & 0.7903 & 0.903 \\
        2 & 0.8749 & $2.44\times 10^{-1}$ & 0.8439 & 0.821 \\
        3 & 0.9420 & $1.65\times 10^{-1}$ & 0.9093 & 0.762 \\
        4 & 0.9948 & $5.16\times 10^{-2}$ & 0.9769 & 0.701 \\
        \hline
    \end{tabular}
    \caption{Dependence of the prepared distribution on the number $m$ of precision qubits used in the QPE filter. The table reports the classical fidelity, total variation distance, basin mass, and phase $0$ acceptance probability for the penalised quantum Metropolis-Hastings algorithm, computed from the exact postselected statevector probabilities.}
    \label{tab:metrics_vs_m}
\end{table}

Table \ref{tab:metrics_vs_m} shows a clear and systematic improvement in all metrics as the number of precision qubits $m$ is increased, as quantified by the classical fidelity, total variation distance, and basin mass defined in Appendix \ref{app:metrics}. The classical fidelity grows monotonically, while the total variation distance decreases by more than a factor of five between $m=1$ and $m=4$, indicating convergence towards the target stationary distribution.

A more refined picture is obtained by examining the basin mass $P_{\mathrm{basin}}(p_X)$, defined as the total probability weight contained in the immediate neighbourhoods of the two energy minima. Already for moderate values of $m$, the local structure within each basin is captured accurately. The dominant source of error instead arises from an imbalance in the relative weights assigned to the two basins. This is reflected in the gradual convergence of $P_{\mathrm{basin}}(p_X)$ towards the analytical value $P_{\mathrm{basin}}(\pi)=0.9815$ as $m$ increases.

The acceptance probability decreases with increasing $m$, as expected, since higher phase resolution sharpens the energy filtering and leads to more selective acceptance. Importantly, this reduction does not compromise the overall quality of the prepared distribution; on the contrary, it correlates with improved agreement with the target distribution.

\begin{figure}[h]
    \centering
    \includegraphics[width=\linewidth]{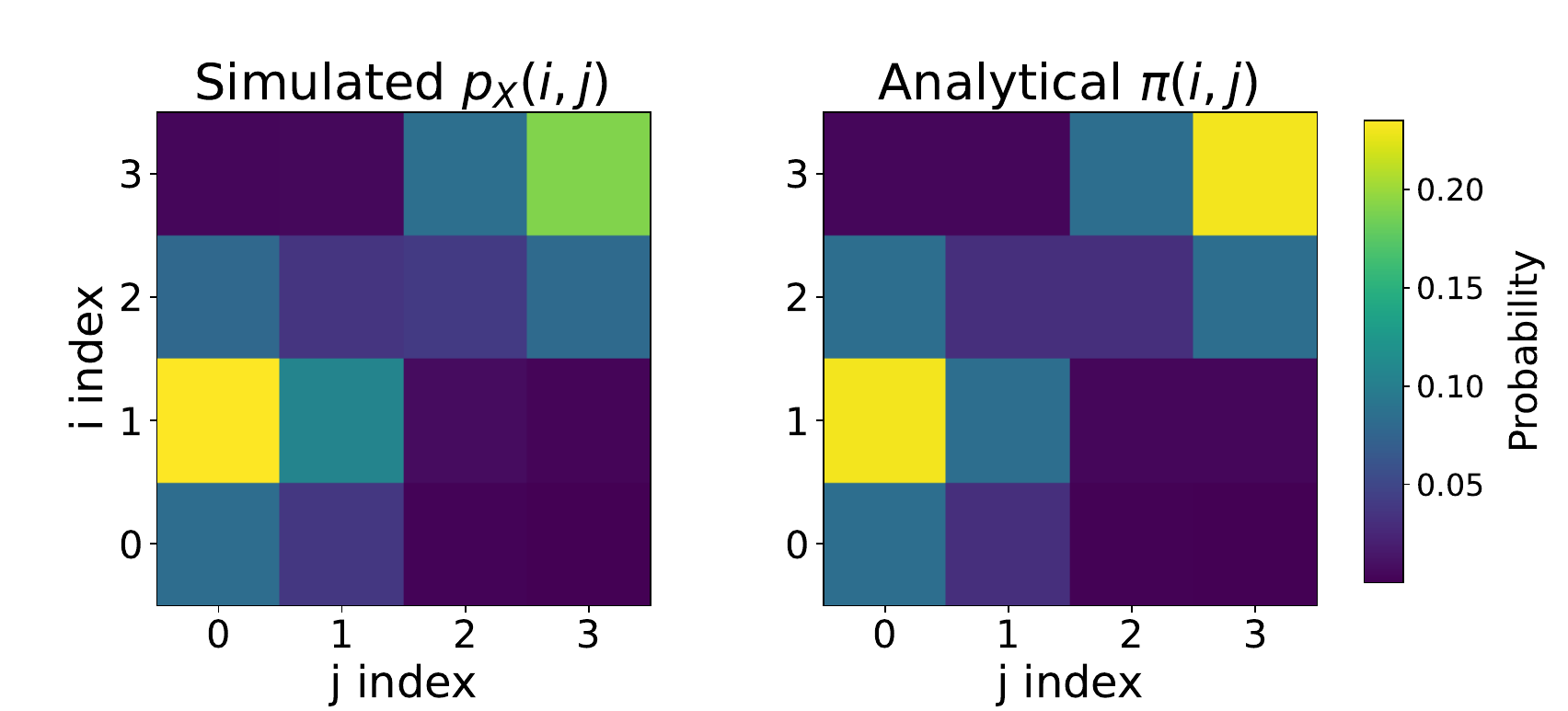}
    \caption{Stationary distribution prepared by the penalised quantum Metropolis-Hastings algorithm with $m=4$ precision qubits (left), compared with the target distribution on the $4\times 4$ grid (right). Both distributions concentrate probability mass around the two low energy basins, with only small residual differences in their relative weights.}
    \label{Distribution V}
\end{figure}

\begin{figure*}[t]
    \centering
    \includegraphics[width=1\linewidth]{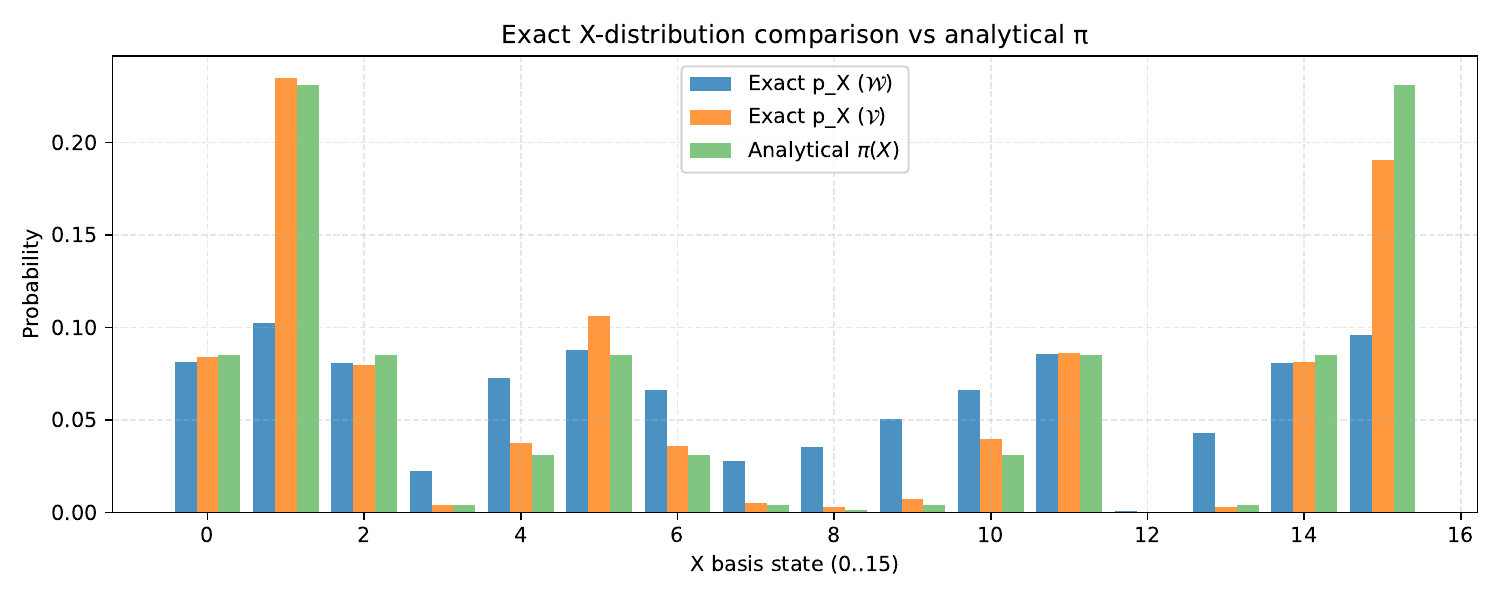}
    \caption{Comparison of the probability distribution over computational basis states $P_X$ obtained from the unmodified qubitized walk with $m=5$ phase ancillas (blue) and from the penalised qubitized walk with $m=4$ phase ancillas (orange), against the analytical target distribution $\pi(X)$ (green). Despite its higher phase precision, the unmodified walk exhibits a substantial misallocation of probability mass, while the penalised walk closely     reproduces the target stationary distribution.}
    \label{comparison}
\end{figure*}

Based on this analysis, we focus on the case $m=4$, which is the largest QPE precision accessible within our classical simulation framework, and examine the corresponding output distribution in more detail below.

FIG. \ref{Distribution V} provides a state by state comparison between the stationary distribution prepared by the penalised quantum Metropolis-Hastings algorithm at $m=4$ and the target distribution on the $4\times 4$ grid. The agreement is visually striking: in both cases, probability mass is sharply localised around the same two energy minima, and the local structure within each basin is faithfully reproduced.

The remaining discrepancy is subtle and highly structured. While the shape and extent of each basin are captured accurately, a small imbalance persists in the relative weights assigned to the two minima. This behaviour is consistent with the finite spectral resolution of the QPE filter at $m=4$, which limits the precision with which small energy differences between metastable regions can be resolved.

Quantitatively, this observation is reflected in a classical fidelity $F(p_X,\pi)=0.9948$ and a total variation distance $d_{TV}(p_X,\pi)=5.2\times10^{-2}$, confirming that the prepared state is globally very close to the target stationary distribution. Inspecting local observables further reveals that the total probability mass contained within a one step neighbourhood of the minima differs from the analytical value by less than $5\times10^{-3}$. The dominant contribution to the remaining total variation distance therefore arises from the exact minima themselves, rather than from distortions of the basin geometry.

Taken together, these results show that already at $m=4$ precision qubits the penalised quantum Metropolis-Hastings algorithm accurately captures both the local and global structure of the target distribution, with residual errors that are well understood and directly attributable to finite phase resolution.

FIG. \ref{comparison} contrasts the output distribution of the penalised quantum Metropolis-Hastings algorithm at $m=4$ with that of the unmodified qubitized walk using a higher phase precision $m=5$. Despite the increased spectral resolution, the unmodified walk exhibits a substantially poorer approximation to the target distribution, as reflected by a classical fidelity of $0.8443$ and a total variation distance of $0.28$. 

A state by state comparison reveals that this degradation is not due to local distortions within each basin, but rather to a pronounced misallocation of probability mass between the two wells. In contrast, the penalised walk accurately reproduces both the localisation of probability mass around the low energy basins and their relative weights, even with fewer phase ancillas.

This comparison highlights a central point of the construction: increased phase precision alone is insufficient to enforce the correct stationary behaviour of the walk. The penalty mechanism plays a crucial role by suppressing spurious stationary components and ensuring convergence to the desired Gibbs distribution. Finite phase resolution then primarily limits the accuracy with which relative basin weights are resolved, rather than the qualitative structure of the sampled distribution.

\subsection{Ising model}
\subsubsection{Model and target distribution}

The Ising model is a paradigmatic system in statistical physics and serves as a standard benchmark for sampling algorithms. Its simplicity, combined with temperature dependent correlations, makes it a natural testbed for analysing both classical and quantum Markov chain Monte Carlo methods.

\paragraph{State space and proposal geometry.}

The state space is the $n$-dimensional Hamming cube,
\begin{equation}
    E = \{-1,+1\}^n,
\end{equation}
whose elements $\sigma = (\sigma_1,\dots,\sigma_n)$ represent spin configurations with $\sigma_i \in \{-1,+1\}$. Hence,
\begin{equation}
    |E| = 2^n.
\end{equation}

In our numerical implementation we take $n=4$, so that $|E|=16$.

The proposal kernel acts by single spin flips. Accordingly, two configurations are neighbours if they differ at exactly one site. The proposal graph is therefore the Hamming cube, which defines the transition neighbourhood of the
Metropolis-Hastings chain.

\paragraph{Hamiltonian.}

The physical interactions are defined on a one-dimensional Ising model with open boundary conditions. The Hamiltonian is
\begin{equation}
    H(\sigma)
    =
    -J \sum_{i=1}^{n-1} \sigma_i \sigma_{i+1}
    - h \sum_{i=1}^{n} \sigma_i,
\end{equation}
where $J$ is the nearest neighbour coupling strength and $h$ is an external magnetic field.

Throughout this section we fix
\[
J = 1,
\qquad
h = 0,
\]
so that
\begin{equation}
    H(\sigma)
    =
    - \sum_{i=1}^{n-1} \sigma_i \sigma_{i+1}.
    \label{Ising_Hamiltonian}
\end{equation}

\paragraph{Target distribution.}

At inverse temperature $\beta > 0$, the target distribution is the Gibbs distribution
\begin{equation}
    \pi_\beta(\sigma)
    =
    \frac{1}{Z_\beta}
    \exp\!\left(-\beta H(\sigma)\right),
\end{equation}
where the partition function is
\begin{equation}
    Z_\beta
    =
    \sum_{\sigma \in E}
    \exp\!\left(-\beta H(\sigma)\right).
\end{equation}

\subsubsection{Performance versus spectral gap}

Under the above choices, the inverse temperature $\beta$ is the sole parameter controlling the structure of the Gibbs distribution. For small $\beta$ (high temperature), $\pi_\beta$ is close to uniform, and the associated Metropolis-Hastings chain on $E$ exhibits a relatively large spectral gap $\delta(\beta)$. As $\beta$ increases, correlations strengthen and the spectral gap decreases.

In the unpenalised construction, the qubitized walk operator $W$ has angular gap
\begin{equation}
    \Delta_W(\beta) = \Omega\!\big(\sqrt{\delta(\beta)}\big).
\end{equation}

In our implementation we instead employ a penalised unitary $\mathcal V$, obtained by applying a phase penalty outside the target subspace. The penalty phase is chosen so as to (heuristically) increase the minimal eigenphase separation from the stationary eigenphase, namely
\begin{equation}
    \Delta_{\mathcal V}(\beta) \gtrsim \Delta_W(\beta).
\end{equation}
Since $\Delta_W(\beta)\in \Omega(\sqrt{\delta(\beta)})$, this implies
\begin{equation}
    \Delta_{\mathcal V}(\beta)\in \Omega(\sqrt{\delta(\beta)}),
\end{equation}
so that the quadratic amplification of the classical spectral gap is preserved.

To isolate the $1-$eigenstate via phase estimation, the phase resolution $2^{-m}$ must be smaller than $\Delta_{\mathcal V}(\beta)$. Up to additive constants depending on the filtering tolerance, this requires
\begin{equation}
m(\beta)
\gtrsim
\log_2\!\left(\frac{1}{\Delta_{\mathcal V}(\beta)}\right)
\lesssim
\frac{1}{2}\log_2\!\left(\frac{1}{\delta(\beta)}\right).
\end{equation}
Thus, increasing $\beta$ reduces the classical spectral gap and correspondingly increases the precision required by the quantum filtering procedure.

\begin{figure}[h]
    \centering
    \includegraphics[width=1\linewidth]{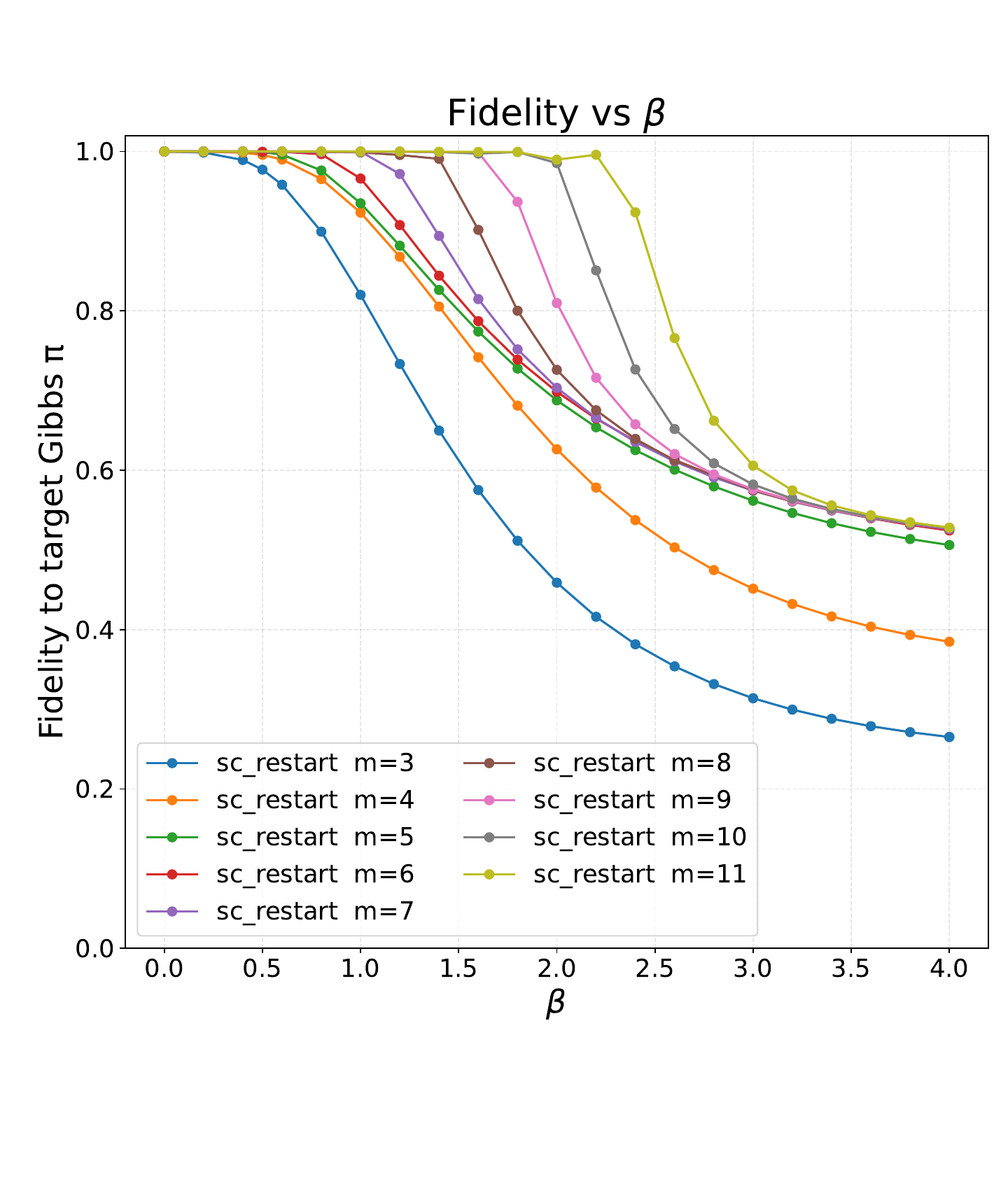}
    \caption{Fidelity between the filtered quantum state and the target Gibbs state as a function of the inverse temperature $\beta$ for $m=3$ to $11$ phase estimation qubits. As $m$ increases, the filtered state more accurately captures the target distribution at larger values of $\beta$. At large $\beta$, the fidelity exhibits an apparent saturation, consistent with the finite size behaviour of the $n=4$ system.} 
    \label{fig:fidelity_vs_beta}
\end{figure}

It is important to emphasize that the algorithm does not explore the configuration space through stochastic trajectories, as in classical MCMC. Instead, it applies a spectral filter to a unitary operator whose eigenstructure encodes the Markov dynamics. Increasing the phase resolution progressively suppresses contributions from eigenmodes with nonzero eigenphases, ideally isolating the stationary eigenvector. When the resolution is insufficient, the output state retains contributions from nearby slow modes, i.e., eigenvectors associated with eigenvalues of the Metropolis kernel close to $1$.

In the present Gibbs Metropolis setting, the structure of these slow modes is shaped by the underlying energy landscape. As $\beta$ increases, the Gibbs distribution concentrates on low energy configurations and transitions between distinct low energy basins become increasingly suppressed under single spin flip dynamics. This induces metastable behaviour in the Markov chain and gives rise to relaxation modes with small spectral gaps. Although these modes are not eigenstates of the Hamiltonian, their support is strongly influenced by the low energy sectors that dominate the Gibbs measure.

This behaviour is reflected in FIG. \ref{fig:fidelity_vs_beta}. When the available phase resolution satisfies $2^{-m}\lesssim \Delta_{\mathcal V}(\beta)$, the stationary component is cleanly isolated and the fidelity remains close to one. Beyond this threshold, the contamination from slow modes appears gradually rather than abruptly, reflecting the detailed distribution of eigenphases near zero. Finally, the apparent saturation observed at large $\beta$ is a finite size effect of the $n=4$ system: for $\beta \gtrsim 4$, the Gibbs distribution becomes strongly concentrated on the two ferromagnetic ground states and changes only marginally with further increases in $\beta$.

\subsubsection{Energy estimation}

The obtained distributions can be used to estimate physical observables in direct analogy with classical Monte Carlo methods \cite{NQS1}. Given an observable $O$ and an output probability distribution $p_X$ extracted from the filtered state, we compute expectation values as
\begin{equation}
    \left< O\right>_{p_X}=\sum_{x\in E}p_X(x)O(x).
\end{equation}
In particular, we consider the energy observable defined by the Ising Hamiltonian (eq. \ref{Ising_Hamiltonian}).

 \begin{figure}[h]
    \centering
    \includegraphics[width=1\linewidth]{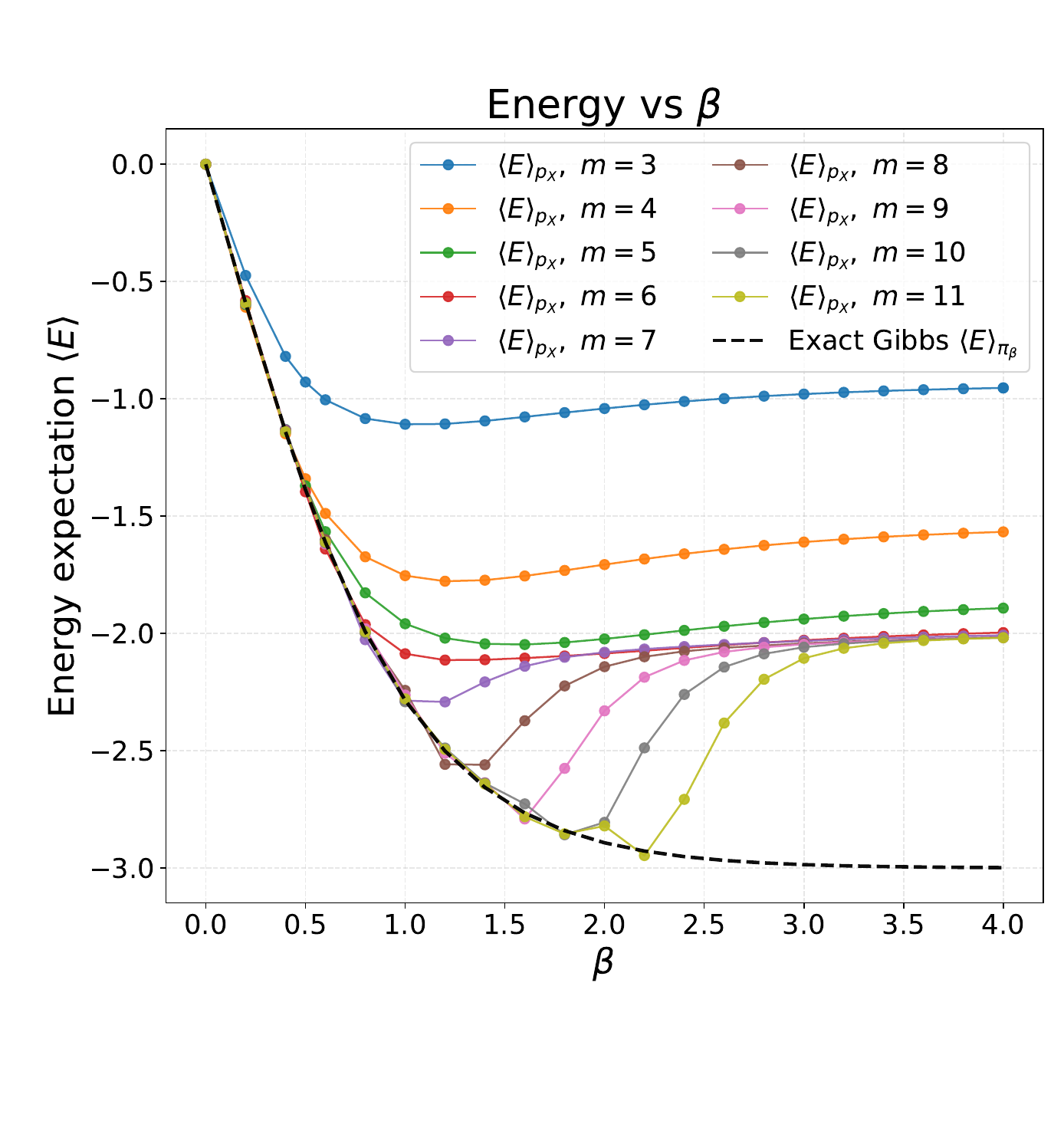}
    \caption{Estimated energy as a function of the inverse temperature $\beta$ for $m=3$ to $11$ phase estimation qubits. Increasing $m$ improves the recovery of the system energy at larger values of $\beta$, while the apparent saturation observed in the large $\beta$ regime reflects the finite size behaviour of the $n=4$ system.}
    \label{fig:energy_vs_beta}
\end{figure}

Figure \ref{fig:energy_vs_beta} shows the estimated energy as a function of $\beta$ for different phase estimation precisions. When the phase resolution satisfies $2^{-m}\lesssim\Delta_{\mathcal V}(\beta)$, the stationary state is accurately isolated and the energy matches the exact Gibbs expectation value. In the under resolved regime, the energy exhibits the same plateau behaviour observed in the fidelity analysis.

A further feature visible in FIG. \ref{fig:energy_vs_beta} is that the difference between consecutive precisions $m$ and $m+1$ in the under resolved regime decreases as $m$ increases. This behaviour follows directly from the finite Fourier resolution of the QPE filter described in Section \ref{sec:level4}. Each additional phase estimation qubit halves the effective width of the Dirichlet kernel centred at eigenphase zero, thereby further suppressing contributions from nonzero eigenphases. 
Once the dominant slow modes are already significantly attenuated, increasing $m$ produces progressively smaller corrections. In this sense, higher phase resolution acts as a higher order spectral refinement, with diminishing incremental impact as the filter becomes sharper.

The structure of the Gibbs Metropolis dynamics further clarifies the observed behaviour. Since the stationary distribution is Gibbs weighted, low energy configurations dominate the measure at large $\beta$. Under single spin flip dynamics, transitions between distinct low energy basins become increasingly suppressed as $\beta$ increases, generating metastable relaxation modes. Although these slow modes are not eigenstates of the Hamiltonian, their support is predominantly concentrated on low energy sectors of configuration space. Consequently, in the under resolved regime the filtered state remains biased toward comparatively low energy configurations even when the full stationary distribution has not yet been isolated.

This mechanism is analysed more explicitly in Appendix \ref{app:Emass Ising}, where we examine the distribution of energy level mass as a function of $\beta$ for different phase resolutions. We observe that increasing the precision of the filter first suppresses contributions from higher energy sectors, while low energy contributions persist until the stationary state is fully isolated. This selective suppression reflects the interplay between Gibbs weighting and the metastable spectral structure of the Markov dynamics.

From a practical perspective, this structured suppression suggests that spectral filtering may preferentially remove high energy contributions even before full convergence to the stationary distribution is achieved. Whether and when this constitutes an advantage compared to variational approaches is problem dependent; in our small scale benchmark it primarily serves as a diagnostic of how finite phase resolution biases the output distribution.

\subsubsection{Basin structure}
In realistic implementations, resource limitations may prevent the algorithm from operating in the fully resolved regime where we have enough precision ancillas to filter the target distribution. Although in the present work this limitation arises from classical simulation constraints, similar restrictions are expected in larger scale implementations. It is thus instructive to examine the qualitative structure of the output distribution in the under resolved regime.

In high-dimensional or multimodal problems, classical Metropolis-Hastings algorithms with local proposals may require a substantial number of steps to transition between weakly connected basins of configuration space. In such settings, the choice of initialization can significantly influence convergence time, and informed initial states are often used in practice to reduce equilibration costs. It is therefore natural to ask whether the quantum workflow exhibits structural features in the under resolved regime that could play a similar role.

When the spectral filter does not fully isolate the stationary eigenstate, the resulting state is not arbitrary; rather, it is dominated by the stationary mode together with the first few low lying spectral modes. 
As discussed in the previous subsection, the contamination is spectrally structured and primarily associated with the first excited mode of the Markov operator. This raises the question of how such structured deviations manifest at the level of collective observables that characterise basin behaviour.

To probe this, we consider two global observables in the Ising model: the magnetization and the number of domain walls.

In the case of the magnetization, it is recovered with numerical precision across the entire range of $\beta$, including the under resolved regime, with an absolute error always below $10^{-7}$. This indicates that, despite incomplete spectral resolution, the quantum prepared distribution preserves the correct magnetization. 

\begin{figure}[h]
    \centering
    \includegraphics[width=1\linewidth]{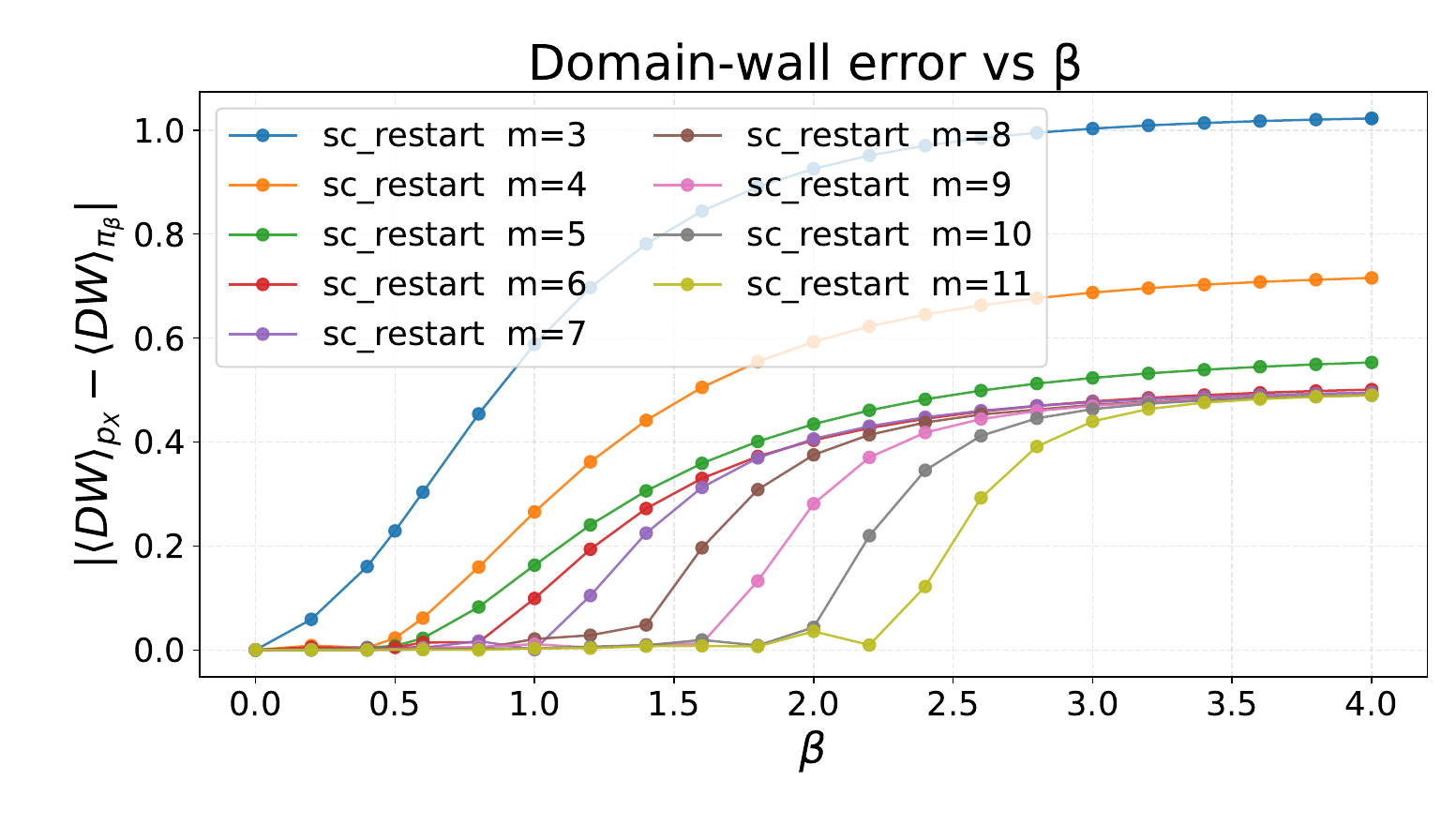}
    \caption{Absolute error in the number of domain walls as a function of the inverse temperature $\beta$ for $m=3$ to $11$ phase estimation qubits. Increasing $m$ allows the algorithm to capture the correct behaviour of the observable at larger values of $\beta$, while the apparent saturation in the large $\beta$ regime reflects the finite size behaviour of the $n=4$ system.}
    \label{fig:dw_error_vs_beta}
\end{figure}

In contrast, FIG. \ref{fig:dw_error_vs_beta} shows that the number of domain walls exhibits a noticeable deviation in the under resolved regime. The error increases as the spectral gap decreases and eventually reaches a plateau, mirroring the behaviour observed in the fidelity and energy estimates. This confirms that not all collective observables are equally robust to finite spectral resolution.

Taken together, these results indicate that the under resolved output state retains nontrivial structural information about the dominant basins of configuration space, even when it does not faithfully reproduce the full stationary distribution. In particular, global order parameters such as magnetization may already be accurately captured at relatively low phase resolution, whereas more sensitive observables reflect the residual spectral bias.

While the under resolved regime does not yield a correct sampler for $\pi_\beta$, the structured nature of the output suggests that it may serve as a meaningful initializer for subsequent refinement procedures. Rather than starting from a completely uninformed distribution, one could begin from a state that already encodes coarse grained basin structure and dominant low lying modes of the dynamics.

\section{\label{sec:level6} Conclusions}
In this work, we have constructed and simulated an explicit quantum implementation of the Metropolis-Hastings algorithm within the framework introduced by Claudon et al. Starting from the classical Markov chain formulation, we analysed the dual representation on the edge space and its realisation as a qubitized quantum walk with a unique stationary eigenstate encoding the target distribution. Beyond reviewing the underlying theory, we focused on the concrete algorithmic and circuit level requirements necessary to make this construction operational.

A central outcome of this study is the identification and resolution of practical obstacles arising in the extraction of the stationary state. We showed that, although the desired quantum state is characterised as the unique $1-$eigenstate within the image of a partial isometry, this property cannot be enforced through consecutive filtering operations in a quantum circuit due to non-commutativity. To address this issue, we introduced and analysed a penalised qubitized walk operator that breaks the global degeneracy of the eigenvalue $1$ while preserving the target state. This penalty mechanism enables a robust filtering procedure based on Quantum Phase Estimation with finite precision and allows a faithful implementation of the algorithm. Beyond enabling the extraction of the stationary state, this construction also allowed us to study in detail the behaviour of the spectral filter generated by quantum phase estimation. In particular, we analysed how the filter behaves when the available phase resolution is insufficient to fully resolve the spectrum, revealing a structured regime in which the resulting quantum dynamics still captures information about basin structure and metastable regions.

We validated these ideas by simulating the full algorithm on two representative systems: a two-dimensional double well potential with $16$ states and a four spin Ising chain. These examples allowed us not only to verify the correctness of the implementation but also to analyse the behaviour of the algorithm under varying spectral conditions. The numerical results demonstrate that the penalised quantum Metropolis-Hastings algorithm prepares states that closely match the target distribution, even when using a limited number of phase ancilla qubits. In contrast, increasing the phase resolution alone without penalisation leads to substantial deviations from the correct stationary distribution. This clearly shows that resolving spectral degeneracies is essential, and that improved filter precision by itself is insufficient. The simulations further reveal how the spectral filtering mechanism governs the exploration of the energy landscape, providing insight into the algorithm’s behaviour as a Gibbs-state sampler when the spectral gap becomes small and the filtering precision is limited.

From a practical perspective, the simulations also make clear that this algorithm is not suited for execution on current noisy intermediate scale quantum (NISQ) devices. The depth of the circuit, dominated by repeated applications of the qubitized walk and the use of the Quantum Fourier Transform, requires a fault-tolerant quantum computer for reliable execution. Moreover, the potential advantage of the quantum Metropolis-Hastings algorithm is expected to emerge only when applied to complex problems with large, structured state spaces, where classical sampling methods suffer from slow mixing and metastability.

Our work demonstrates the feasibility of implementing a quantum version of the Metropolis-Hastings algorithm in a realistic circuit model, clarifies the resources required for its successful execution, and provides guidance on how quantum sampling techniques may be integrated into broader computational workflows in the future. An interesting direction for further research is the exploration of hybrid quantum-classical strategies. In particular, the use of quantum Metropolis-Hastings algorithm as an initialisation procedure to prepare high quality samples in challenging, multimodal energy landscapes, after which a classical Markov chain Monte Carlo method running on high performance computing infrastructure could be used for local refinement. Such a hybrid approach could leverage the strengths of quantum sampling in overcoming global energy barriers while retaining the efficiency of classical methods for fine tuning and local exploration.

\section*{Acknowledgments}
The authors acknowledge funding from the Spanish Ministry for Digital Transformation and of Civil Service of the Spanish Government through the QUANTUM ENIA project call - Quantum Spain, EU through the Recovery, Transformation and Resilience Plan – NextGenerationEU within the framework of the Digital Spain 2026. The work has also been supported by the projects CEX2021-001148-S, and PID2023-147979NB-C21 from the MCIN/AEI and MICIU/AEI /10.13039/501100011033 and by FEDER, UE, by the Departament de Recerca i Universitats de la Generalitat de Catalunya, research group MPiEDist (2021 SGR 00412).

In addition, the authors received financial support from the European Union, through the EuroHPC Joint Undertaking and its members under the Grant Agreement Nº 101159808, including top-up funding by the Ministry for Digital Transformation and the Civil Service of the Spanish Government.

\bibliographystyle{unsrt}
\bibliography{biblio}

\begin{thebibliography}{10}

\bibitem{RadiationTransport1}
E.W. Larsen.
\newblock An overview of neutron transport problems and simulation techniques.
\newblock In {\em Computational Methods in Transport}, pages 513--534, 2006.

\bibitem{MonteCarloMethodsBook}
M.H. Kalos and P.A Whitlock.
\newblock {\em Monte Carlo Methods}.
\newblock WILEY-VCH Verlag GmbH \& Co. KGaA, second edition, 2008.

\bibitem{NQS1}
K.~Choo, A.~Mezzacapo, and G.~Carleo.
\newblock Fermionic neural-network states for ab-initio electronic structure.
\newblock {\em Nature Communications}, 11(1), 2020.

\bibitem{NQS2}
J.~W.~T. Keeble, M.~Drissi, A.~Rojo-Francàs, B.~Juliá-Díaz, and A.~Rios.
\newblock Machine learning one-dimensional spinless trapped fermionic systems
  with neural-network quantum states.
\newblock {\em Physical Review A}, 108(6), 2023.

\bibitem{InverseProblems1}
J.P. Kaipio and E.~Somersalo.
\newblock {\em Statistical and Computational Inverse Problems}.
\newblock Springer, first edition, 2004.

\bibitem{InverseProblems2}
A.~Beskos, A.~Jasra, E.A. Muzaffer, and A.M. Stuart.
\newblock Sequential monte carlo methods for bayesian elliptic inverse
  problems.
\newblock {\em Statistics and Computing}, 25, 2015.

\bibitem{MCMCProblems1}
G.~Roberts and J.~Rosenthal.
\newblock General state space markov chains and mcmc algorithms.
\newblock {\em Probability Surveys}, 1, April 2004.

\bibitem{MCMCProblems2}
K.~Wu, S.~Schmidler, and Y.~Chen.
\newblock Minimax mixing time of the metropolis-adjusted langevin algorithm for
  log-concave sampling.
\newblock {\em Journal of Machine Learning Research}, 23, 2022.

\bibitem{MCMCProblems3}
A.~Beskos, M.~Girolami, S.~Lan, P.~Farrell, and A.~Stuart.
\newblock Geometric mcmc for infinite-dimensional inverse problems.
\newblock {\em J. Comput. Phys.}, 335, 2016.

\bibitem{MCMCProblems4}
Q.~Zhou and H.~Chang.
\newblock Complexity analysis of bayesian learning of high-dimensional dag
  models and their equivalence classes.
\newblock {\em The Annals of Statistics}, 51, 2021.

\bibitem{Szegedy}
M.~Szegedy.
\newblock Quantum speed-up of markov chain based algorithms.
\newblock In {\em 45th Annual IEEE Symposium on Foundations of Computer
  Science}, pages 32--41, 2004.

\bibitem{markovchainslevin}
D.~Levin and Y.~Peres.
\newblock {\em Markov Chains and Mixing Times}.
\newblock American Mathematical Society, second edition, 2009.

\bibitem{Lemieux}
J.~Lemieux, B.~Heim, D.~Poulin, K.~Svore, and M.~Troyer.
\newblock Efficient quantum walk circuits for metropolis-hastings algorithm.
\newblock {\em Quantum}, 4, June 2020.

\bibitem{Temme}
K.~Temme, T.J. Osborne, K.G.H. Vollbrecht, D.~Poulin, and F.~Verstraete.
\newblock Quantum metropolis sampling.
\newblock {\em Nature}, 471(7336), 2011.

\bibitem{Montanaro}
A.~Montanaro.
\newblock Quantum speedup of monte carlo methods.
\newblock {\em Proceedings of the Royal Society A: Mathematical, Physical and
  Engineering Sciences}, 471(2181), 2015.

\bibitem{Claudon}
B.~Claudon, P.~Rodenas-Ruiz, J.P. Piquemal, and P.~Monmarché.
\newblock Quantum circuits for the metropolis-hastings algorithm, 2025.

\bibitem{firstcourseMCMC}
D.S. Alonso and O.~Al-Ghattas.
\newblock A first course in monte carlo methods, 2024.

\bibitem{Generalized_QSVT}
C.~Sünderhauf.
\newblock Generalized quantum singular value transformation, 2023.

\bibitem{discriminant}
F.~Magniez, A.~Nayak, J.~Roland, and M.~Santha.
\newblock Search via quantum walk.
\newblock {\em SIAM Journal on Computing}, 40(1), 2011.

\bibitem{QPE}
R.~Cleve, A.~Ekert, C.~Macchiavello, and M.~Mosca.
\newblock Quantum algorithms revisited.
\newblock {\em Proceedings of the Royal Society of London. Series A:
  Mathematical, Physical and Engineering Sciences}, 454(1969), 1998.

\bibitem{Unification}
J.M. Martyn, Z.M. Rossi, A.K. Tan, and I.L. Chuang.
\newblock Grand unification of quantum algorithms.
\newblock {\em PRX Quantum}, 2(4), 2021.

\bibitem{DirichletKernel}
A.M. Bruckner, J.B. Bruckner, and B.S. Thomson.
\newblock {\em Real Analysis}.
\newblock Prentice-Hall, first edition, 1997.

\bibitem{EfficientFactorization}
S.~Parker and M.B. Plenio.
\newblock Efficient factorization with a single pure qubit and logn mixed
  qubits.
\newblock {\em Physical Review Letters}, 85(14), 2000.

\bibitem{github_repo}
M.~Carrasco-Arango.
\newblock Quantum metropolis hastings: Circuit implementation, 2026.
\newblock \url{https://github.com/bsc-wdc/q-mcmc}.

\end{thebibliography}

\appendix
\section{Circuit implementations}
\label{app:circuits}

In this appendix we present the quantum circuits implementing the penalised version of the quantum Metropolis-Hastings algorithm. The full implementation, including circuit generation and simulation code, is available in the accompanying repository \cite{github_repo}.

\begin{figure*}[t]
    \centering
    \includegraphics[width=0.99\linewidth]{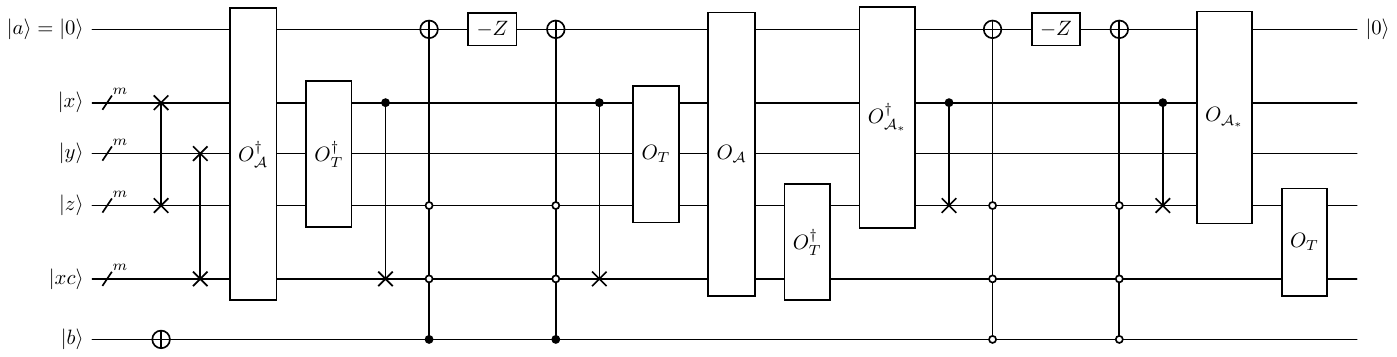}
    \caption{Quantum circuit implementing the qubitized walk operator $\mathcal{W}$.}
    \label{W circuit}
\end{figure*}

FIG. \ref{W circuit} shows the quantum circuit that implements the qubitized walk operator $\mathcal{W}$ introduced in \cite{Claudon}. The operator acts exclusively on the system qubits. Following the notation of Theorem \ref{teorema}, the circuit begins by applying a Pauli $X$ gate to the ancilla $b$, followed by two SWAP operations exchanging the register pairs $(x, z)$ and $(y, xc)$.

The core of the construction is a reflection about the image of the isometry $\boxtimes$, which can be decomposed as the product of two reflections associated with the isometries $\square$ and $\square_*$. Each reflection is implemented using a controlled Pauli $Z$ gate together with a PUE. In this process, the state of the register $x$ is copied into the register $xc$ (or into $z$ in the case of $\square_*$), after which the proposal and acceptance oracles are applied.

This procedure does not violate the no cloning theorem, since the `copy' operation is only required to act correctly on the range of $\boxtimes$, where the register $xc$ is initialised to $\ket{0}$, allowing the use of CNOT gates to mimic this transformation.

We now describe the construction of the proposal and acceptance oracles used in the implementation of $\mathcal{W}$.

\subsection{\texorpdfstring{Proposal oracle $O_T$}{Proposal oracle OT}}

The proposal oracle $O_T$ prepares, for each current state $x$, a coherent superposition over neighbouring states $y$ according to the chosen proposal kernel $T(x,y)$. Its implementation is topology dependent: the circuit structure encodes the adjacency relation of the proposal graph and produces a clean coin register that is uncomputed at the end of the operation.

\subsubsection{Bidimensional double well (periodic grid)}

To implement the random proposal step, the topology of the proposal graph must be taken into account. In the present case, the state space is a two-dimensional grid in which each vertex has four neighbours. This allows the proposal to be implemented using a two-qubit coin register encoding the direction of the move along the vertical or horizontal axis.

\begin{figure}[h!]
    \centering
    \includegraphics[width=\linewidth]{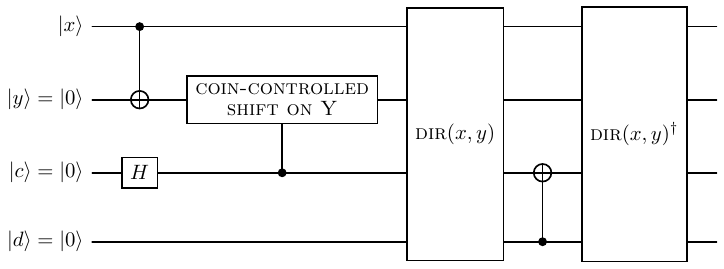}
    \caption{Quantum circuit implementing the proposal oracle $O_T$ for the $4\times 4$ bidimensional grid with periodic nearest neighbour connectivity.}
    \label{fig:OT_doublewell_circuit}
\end{figure}

Before describing the circuit in detail, we specify how classical states are encoded into quantum registers. Each state $x\in E$ is identified with a pair $(i,j)$, where $i$ and $j$ denote the horizontal and vertical coordinates on the grid, respectively. This representation is encoded using four computational qubits as

\begin{equation}
    \ket{x}
    \;=\;
    \ket{i,j}
    \;=\;
    \ket{i_0, i_1, j_0, j_1},
\end{equation}
where
\begin{equation}
    i = i_0 + 2 i_1, \qquad
    j = j_0 + 2 j_1, \qquad
    i_0,i_1,j_0,j_1 \in \{0,1\}.
\end{equation}

As shown in FIG. \ref{fig:OT_doublewell_circuit}, the proposal step begins by copying the content of the register $x$ into the register $y$. This acts as a copy on the subspace where $y$ is initialised in $\ket{0}$ (the only relevant subspace for the workflow), and can therefore be implemented using CNOT gates.

A two qubit coin register $c$ is then used to encode one of the four possible directions of motion: $\ket{00}$ corresponds to a downward move $(i' = i+1 \bmod 4)$, $\ket{01}$ to an upward move $(i' = i-1 \bmod 4)$, $\ket{10}$ to a rightward move $(j' = j+1 \bmod 4)$, and $\ket{11}$ to a leftward move $(j' = j-1 \bmod 4)$.

To uncompute the coin register, the direction of the proposed move is recomputed from the data registers $(x,y)$ into an auxiliary scratch register $d$, without accessing the original coin register $c$. The circuit implementing the function $d(x,y)$ is composed entirely of reversible gates, including Pauli $X$ gates to realise open controls and multi-controlled NOT (MCX) gates, and does not require additional ancillas.

At this stage, the registers $c$ and $d$ encode the same information. A CNOT gate controlled by $d$ is therefore applied to reset the coin register $c$ to $\ket{0}$. Finally, the inverse of the circuit computing $d(x,y)$ is applied, returning both auxiliary registers $c$ and $d$ to the state $\ket{0}$ and completing the uncomputation.

\subsubsection{Ising model (single-spin flip)}

The proposal oracle for the Ising model follows the same general structure as in the bidimensional double well case. The proposal graph is now the Hamming cube, where two configurations are neighbours if they differ at exactly one spin. For $n=4$, each configuration therefore has four neighbours, corresponding to flipping one of the four spins.

As in the previous construction, the proposal step begins by copying the content of the register $x$ into the register $y$. Since the register $y$ is initialised in the state $\ket{0}$, this copy operation can be implemented using CNOT gates and acts correctly on the only subspace relevant for the workflow.

A two qubit coin register $c$ is then prepared in a uniform superposition over the four computational basis states, encoding the index $k$ of the spin to be flipped. The proposal update is implemented by applying a Pauli $X$ gate to the qubit $y_k$, controlled on the coin register being in the state $\ket{k}$. Operationally, this is realised using pattern-controlled multi-controlled NOT (MCX) gates.

In contrast to the bidimensional double well construction, no auxiliary scratch register is required to uncompute the coin. Since the proposal guarantees that $y$ differs from $x$ in exactly one spin, the identity of the flipped spin can be recomputed directly from the pair of registers $(x,y)$. The uncomputation proceeds by applying CNOT gates that accumulate the parities $x_j \oplus y_j$ into the coin qubits, thereby resetting them to the state $\ket{0}$. The entire procedure is reversible and leaves all ancillas clean.

This direct uncomputation strategy reduces the ancillary qubit overhead compared to the double well implementation, where an explicit scratch register is required to reconstruct the direction of motion.

\subsection{\texorpdfstring{Acceptance oracle $O_\mathcal{A}$}{Acceptance oracle OA}}
\begin{figure}[h]
    \centering
    \includegraphics[width=1\linewidth]{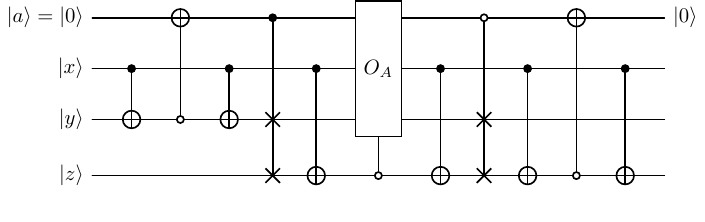}
    \caption{Quantum circuit implementing the acceptance operator of $\mathcal{P}$, $O_\mathcal{A}$.}
    \label{O mathcal A}
\end{figure}

\begin{figure}
    \centering
    \includegraphics[width=0.6\linewidth]{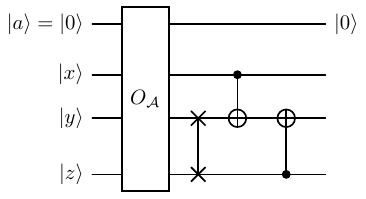}
    \caption{Quantum circuit implementing the acceptance operator of $\mathcal{P}^*$, $O_\mathcal{A^*}$.}
    \label{O mathcal A ast}
\end{figure}

The circuits implementing the acceptance oracle, shown in FIG. \ref{O mathcal A} and FIG. \ref{O mathcal A ast}, are constructed following the procedure introduced in \cite{Claudon}. They are composed of CNOT and SWAP gates together with a controlled implementation of the classical Metropolis-Hastings acceptance oracle, realised as a controlled rotation acting on the ancilla $a$. The rotation depends on the values of the registers $x$ and $y$ and is guarded by the equality condition $x=z$.

When the condition $x=z$ is satisfied, a rotation by an angle $\phi(x,y)$ is applied to the ancilla $a$ according to
\begin{align}
    &\ket{0}_a
    \;\longmapsto\;
    \sqrt{1 - A(x,y)}\ket{0}_a
    + \sqrt{A(x,y)}\ket{1}_a,
    \\[4pt]
    &\ket{1}_a
    \;\longmapsto\;
    -\,\sqrt{A(x,y)}\ket{0}_a
    +  \sqrt{1 - A(x,y)}\ket{1}_a.
\end{align}
On the ancilla subspace spanned by $\{\ket{0},\ket{1}\}$, this transformation corresponds exactly to a rotation $R_y(\phi(x,y))$, where the angle is chosen such that
\begin{equation}
    \phi(x,y)=2\arcsin \left(\sqrt{A(x,y)}\right).
\end{equation}
This choice ensures that the rotation encodes the classical Metropolis-Hastings acceptance probability $A(x,y)$ (see Eq. \ref{eq:MH acceptance kernel}). If the condition $x\neq z$ is not satisfied, the oracle acts as the identity.

\subsection{\texorpdfstring{Penalised qubitized walk $\mathcal{V}$}{Penalised qubitized walk V}}

Given the proposal and acceptance oracles, we can construct the PUEs required to implement the qubitized walk $\mathcal{W}$. As discussed in the main text, however, $\mathcal{W}$ is not used directly. Instead, we work with a penalised version $\mathcal{V}=\left(\Pi_\boxtimes+e^{i\varphi}(Id-\Pi_\boxtimes)\right)\mathcal{W}$, as introduced in section \ref{sec:level4}. 

Since projectors cannot be implemented directly in a quantum circuit, the operator $\Pi_{\boxtimes}$ must be realised implicitly through a coherent membership test. This is achieved by exploiting the defining property of the isometry $\boxtimes$: a state belongs to $\mathrm{Im}(\boxtimes)$ if applying the adjoint isometry $\boxtimes^\dagger$ maps it to a state in which the auxiliary registers $z$, $xc$, $a$, $c$, and $d$ are all in the computational state $\ket{0}$.

Operationally, this condition is checked coherently using a set of CNOT gates together with a dedicated penalty flag qubit, which is flipped if and only if the state lies in $\mathrm{Im}(\boxtimes)$. Conditioned on the value of this flag, a phase $e^{i\varphi}$ is applied to all states orthogonal to $\mathrm{Im}(\boxtimes)$. Finally, the computation is uncomputed by reapplying $\boxtimes$, restoring the original register structure.

This unitary penalisation procedure effectively realises the action of the operator $\Pi_{\boxtimes} + e^{i\varphi}(Id -\Pi_{\boxtimes})$ without ever explicitly implementing a projection. The resulting operator $\mathcal{V}$ is then used wherever the qubitized walk is required, in particular within the Quantum Phase Estimation routine employed for eigenstate filtering.

\subsection{Quantum Phase Estimation}
\label{sec:qpe}
\begin{figure}[h]
    \centering
    \includegraphics[width=1\linewidth]{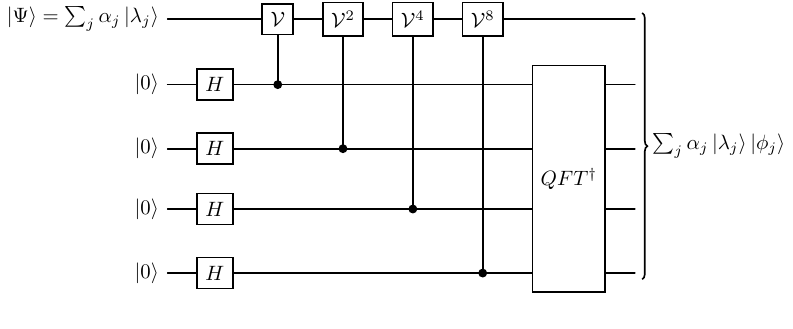}
    \caption{Quantum circuit implementing the Quantum Phase Estimation (QPE) algorithm.}
    \label{QPE circuit}
\end{figure}

The eigenstate filtering step is implemented using the Quantum Phase Estimation algorithm, whose circuit is shown in Fig. \ref{QPE circuit}. The QPE is applied to the penalised qubitized walk operator $\mathcal{V}$. The role of QPE in this context is not to estimate eigenvalues per se, but to coherently filter the eigenstates associated with eigenphase zero, corresponding to eigenvalue $1$.

If $\ket{\lambda}$ is an eigenstate of $\mathcal{V}$ with eigenvalue $e^{i 2\pi \phi}$, QPE performs the transformation $\ket{\lambda}\ket{0} \rightarrow \ket{\lambda}\ket{\tilde{\phi}}$, where $\tilde{\phi}$ is an $m$-bit estimate of the phase $\phi$ stored in the phase register.

The phase estimate is represented in binary as
\begin{equation}
    \phi=0.\phi_1\phi_2\ldots\phi_m \text{ (binary)},
\end{equation}
so that, with $m$ phase register ancillas, the achievable phase resolution scales as $2^{-m}$.

In the subsequent postselection step, only those outcomes for which the phase register is measured in the state $\ket{0}$ are retained. This implements a finite-precision spectral filter that approximately projects onto the $1$-eigenspace of $\mathcal{V}$.

\subsection{Dual to vertex decoding transformation}

\begin{figure}[h!]
    \centering
    \includegraphics[width=1\linewidth]{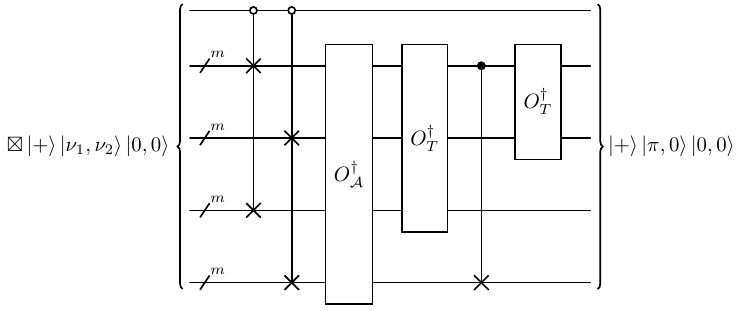}
    \caption{Quantum circuit implementing the final transformation on the ancilla $b$ and the four main registers.}
    \label{Final transformation circuit}
\end{figure}

After the QPE based filtering step, the resulting quantum state is close to the $1-$eigenstate of the penalised qubitized walk operator $\mathcal{V}$. When the penalisation is correctly implemented, this target state is uniquely characterised by simultaneously lying in the image of the partial isometry $\boxtimes$ and in the $1-$eigenspace of the unpenalised qubitized walk operator $\mathcal{W}$. This state encodes the dual stationary distribution $\nu$ over the directed edges of the proposal graph.

The purpose of the final transformation is to map this edge based encoding into a quantum state whose amplitudes in the register $x$ reproduce the stationary distribution $\pi$ of the original classical Metropolis-Hastings chain. This transformation is unitary and acts only on the system registers and the ancilla $b$, preserving coherence throughout the process.

Operationally, the circuit shown in Fig. \ref{Final transformation circuit} performs a sequence of controlled SWAP operations conditioned on the ancilla $b$, followed by the application of a projected unitary encoding and a final proposal oracle. These steps effectively undo the edge encoding introduced by the dual construction, marginalising over the head register while coherently accumulating the correct probability weights on the tail register.

As a result, the filtered state $\boxtimes\ket{+}\ket{\nu}\ket{0}$ is transformed into a state of the form $\ket{+}\ket{\pi,0}\ket{0}$, where the register $x$ encodes the stationary distribution $\pi$. Measuring this register in the computational basis therefore yields samples distributed according to the target classical Metropolis–Hastings distribution.

\section{Evaluation metrics}
\label{app:metrics}

In this appendix we define the metrics used to quantitatively compare the probability distribution prepared by the quantum Metropolis-Hastings algorithm with the target stationary distribution. All metrics considered here are classical distances between probability distributions on the finite state space $E$ and are computed from measurement outcomes of the quantum circuit.

Let $\ket{\Psi_f}$ denote the final quantum state obtained after the QPE based filtering and the final transformation described in Section \ref{sec:level4}. Measuring the register $x$ in the computational basis yields samples $x \in E$ distributed according to a probability distribution $p_X$.

In practice, $p_X$ is estimated either from repeated projective measurements (shots) or, in the case of statevector simulation, from the exact measurement probabilities of $\ket{\Psi_f}$. All reported metrics are computed using this empirical distribution $p_X$ and the analytical target distribution $\pi$.

\subsection{Classical fidelity}

To quantify the global overlap between the prepared distribution $p_X$ and the target distribution $\pi$, we use the classical fidelity
\begin{equation}
    F(p_X,\pi) \;=\; \left(\sum_{x\in E} \sqrt{p_X(x)\,\pi(x)}\right)^2.
\end{equation}
This quantity takes values in $[0,1]$, with $F(p_X,\pi)=1$ if and only if $p_X=\pi$. The classical fidelity is symmetric, insensitive to small local fluctuations, and provides a smooth measure of overall agreement between distributions.

\subsection{Total variation distance}

As a complementary metric, we consider the total variation distance
\begin{equation}
    d_{\mathrm{TV}}(p_X,\pi)
    \;=\;
    \frac{1}{2}\sum_{x\in E} \left| p_X(x)-\pi(x) \right|.
\end{equation}
The total variation distance has a clear operational interpretation as the maximum probability of distinguishing the two distributions with a single sample. Unlike the fidelity, it is sensitive to local discrepancies and therefore highlights misallocation of probability mass.

\subsection{Local basin mass}

In order to analyse the structure of deviations between $p_X$ and $\pi$ in Section \ref{sec:level4}, we additionally consider the probability mass assigned to local neighbourhoods of the energy minima. Given a minimum $x^\star=(i^\star,j^\star) \in E$ and a radius $r\ge 0$, we define a local basin as the $L^\infty$ neighbourhood
\begin{equation}
    \mathcal{B}_\infty(x^\star,r)
    \;=\;
    \{(i,j)\in E \mid |i-i^\star|\le r,\ |j-j^\star|\le r\}.
\end{equation}
For a set of minima $\{x^\star\}$, we then define the corresponding basin mass as
\begin{equation}
    P_{\mathrm{basin}}(p_X)
    \;=\;
    \sum_{x \in \bigcup_{x^\star}\mathcal{B}_\infty(x^\star,r)} p_X(x),
\end{equation}
and compare this quantity with the corresponding value computed from the target distribution $\pi$.

In the numerical experiments we report results for $r=1$, which corresponds to the immediate spatial neighbourhood of each minimum. This metric isolates discrepancies in the relative weight assigned to different metastable basins and allows us to distinguish global imbalances between basins from local inaccuracies within each basin. In the double well simulations, this analysis reveals that the dominant contribution to the total variation distance arises from small errors in the relative weights of the two wells rather than from local distortions of their internal structure.

\section{Semiclassical implementation of the QPE filter}
\label{app:semiclassical QPE filter}
\begin{figure}[ht]
    \centering
    \includegraphics[width=1\linewidth]{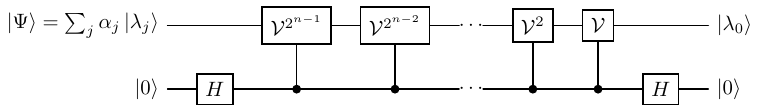}
    \caption{Semiclassical quantum circuit implementing the QPE based filter (with a single recycled phase ancilla).}
    \label{fig: semiclassical QPE}
\end{figure}

In this appendix we describe the semiclassical implementation of the QPE filter that uses a single phase ancilla qubit. The key idea is to reduce circuit width by recycling the phase register, at the cost of additional circuit depth and intermediate measurements with classical feedforward.

This construction follows the semiclassical quantum Fourier transform technique introduced in \cite{EfficientFactorization}. The inverse quantum Fourier transform (iQFT) appearing in QPE is a sequence of gates applied sequentially on the qubits. Consequently, instead of coherently applying the full iQFT on an $m$-qubit phase register and measuring all ancillas at the end, one may measure the phase qubits sequentially. The controlled phase rotations are then replaced by classically conditioned single qubit phase gates acting on the recycled ancilla, where the conditioning depends on the outcomes of the previous measurements. 

\begin{figure*}[ht]
    \centering
    \includegraphics[width=1\linewidth]{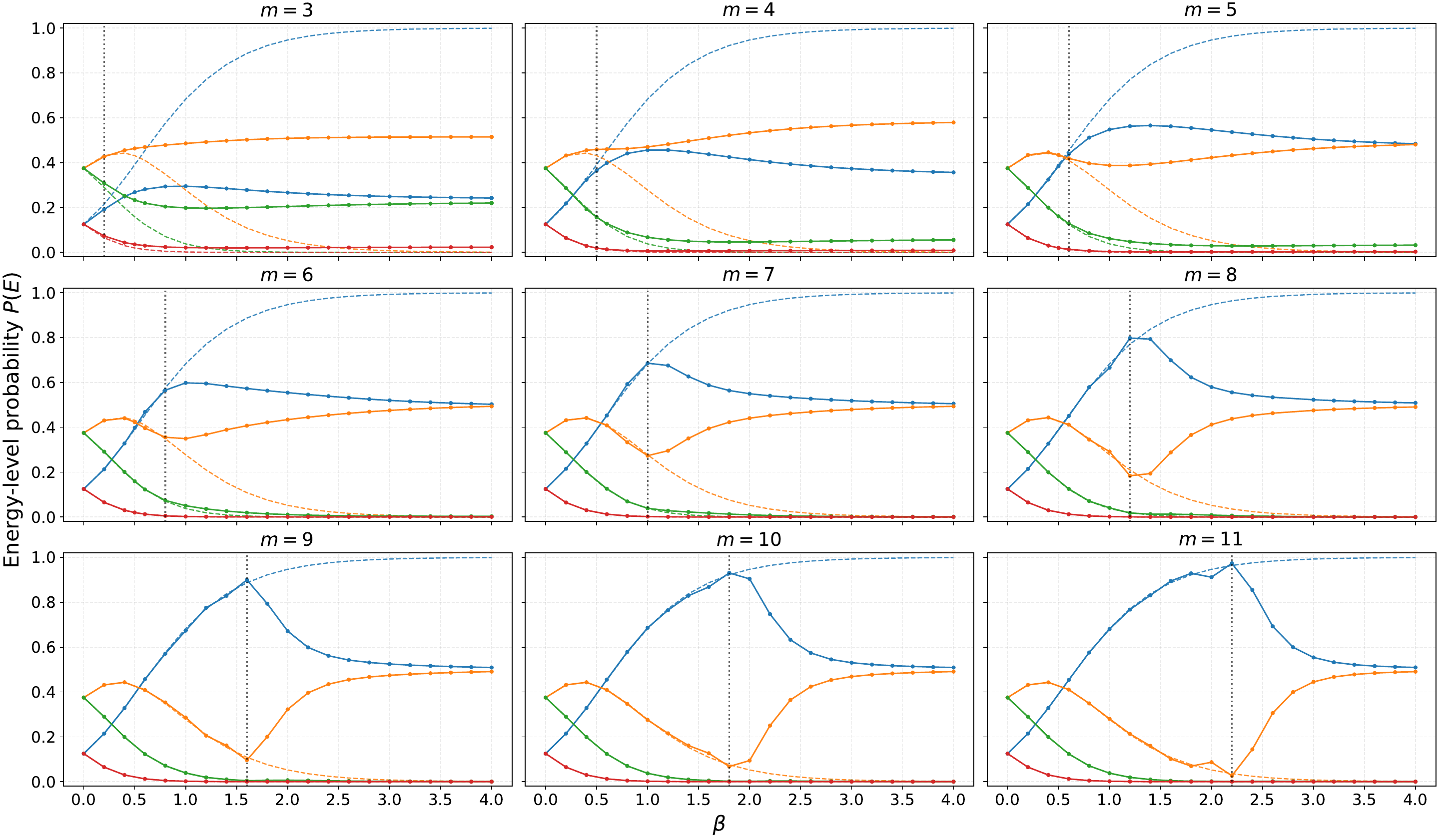}
    \caption{Energy level probability mass of the filtered state as a function of the inverse temperature $\beta$ for increasing QPE precision $m$. Panels correspond to $m=3$ to $m=11$. Solid curves show the probability mass assigned by the filtered distribution to each energy sector, while dashed curves indicate the exact Gibbs probabilities. Colours identify the four energy levels of the $N=4$ Ising Hamiltonian: blue ($E=-3$), orange ($E=-1$), green ($E=1$), and red ($E=3$). The vertical dotted line marks the largest value of $\beta$ for which the fidelity between the filtered state and the exact Gibbs distribution remains above the chosen threshold $(0.992)$.}
    \label{fig:E_mass_grid}
\end{figure*}

In our implementation, the filter is realised by projecting the phase ancilla onto the state $\ket{0}$. Under this postselection, the classically conditioned phase corrections are never activated and can therefore be omitted. Moreover, the two Hadamard gates on the recycled ancilla satisfy $H^2=\mathbb{I}$, which further simplifies the resulting circuit. The final semiclassical QPE filter circuit is shown in FIG. \ref{fig: semiclassical QPE}.

\section{Energy sector probabilities in the Ising model}
\label{app:Emass Ising}

To clarify the structure of the under resolved regime, we analyse how the probability mass of the filtered state is distributed across the different energy levels of the Ising Hamiltonian as a function of the inverse temperature $\beta$ and of the QPE precision $m$.

Configurations are grouped according to their energy $E$, and we compute the total probability weight assigned to each energy sector. This provides a direct view of how the spectral filter reshapes the distribution as its resolution increases.

Figure \ref{fig:E_mass_grid} shows the distribution of probability mass across the energy levels for increasing QPE precision $m$. In the exact Gibbs distribution, the first excited energy level carries the largest probability weight at moderate inverse temperatures ($\beta \lesssim 0.6$), while the ground level becomes dominant as $\beta$ increases further.

At low precision ($m=3$), the filter is unable to resolve the lowest spectral modes. Although the highest excited level is already suppressed, the first excited level remains dominant even at large $\beta$, indicating that the ground state sector is not properly distinguished. Increasing the precision to $m=4$ improves the hierarchy of suppression: the highest excited level becomes negligible and the second excited level begins to decrease significantly. However, the crossover between the first excited and ground levels is still not reproduced.

A qualitative change appears at $m=5$, where the correct crossover is recovered and the ground level becomes the dominant sector at sufficiently large $\beta$. This indicates that the spectral resolution is sufficient to distinguish the two lowest energy sectors. For $m=6$, the behaviour remains qualitatively similar: higher excited levels are essentially suppressed, while the dominant residual contribution arises from the first excited level, whose probability remains close to that of the ground level at large $\beta$.

For $m\geq 6$, the global structure of the curves remains the same. Increasing the QPE precision mainly extends the range of inverse temperatures over which the filtered distribution matches the exact Gibbs probabilities. However, for sufficiently large $\beta$, the filtered state systematically approaches a regime in which the probability mass is shared almost equally between the ground state and the first excited sector. In this large $\beta$ regime the filter cannot fully resolve the energy gap between these two lowest levels, leading to an approximate $50\%$–$50\%$ mixture between them.

Overall, the figure reveals a hierarchical suppression of excited energy sectors as the spectral resolution increases. Higher excited levels disappear first, while the first excited level remains the dominant residual component in the under resolved regime.  

\end{document}